\documentclass[a4paper]{article}

\usepackage{fullpage}
\usepackage{times}

\usepackage{latexsym}
\usepackage{diagrams}
\usepackage{headers}
\usepackage{amssymb}
\usepackage{amsmath}

\usepackage{proof}

\DeclareFontFamily{U}{stmary}{}
\DeclareFontShape{U}{stmary}{}{}{<5> <6> <7> <8> <9> <10> gen * stmary %
  <10.95> <12> <14.4> <17.28> <20.74> <24.88> stmary10}{}
\DeclareSymbolFont{linear}{U}{stmary}{}{}
\DeclareMathSymbol{\por}{\mathbin}{linear}{'117}
\DeclareMathSymbol{\with}{\mathbin}{linear}{'116}

\newcommand{\lolli}{\multimap}

\newcommand{\qed}[0]{\hfill$\blacksquare$}

\newtheorem{define}{Definition} 
 \newtheorem{prop}{Proposition}
\newtheorem{lemma}{Lemma}

\newcommand{\limp}{\lolli}
\newcommand{\bc}{\begin{center}}
\newcommand{\ec}{\end{center}}
\newcommand{\ds}{\displaystyle}

\newcommand{\ov}{\overline}
\newcommand{\bimp}{\mimp}

\def\ie{{\it i.e.},}
\def\eg{{\it e.g.},}
\def\etc{{\it etc.}}

 \newenvironment{proof}%
        {\begin{list}{ {\bf Proof}}%
                           {\setlength{\leftmargin}{\labelwidth}} \item }%
        {\qed \end{list}}

\newcommand{\BI}{{\bf BI}}
\newcommand{\mimp}{\,{-\!\!*}\,}

\newcommand{\LBI}{\mbox{\bf LBI}}
\newcommand{\side}{\mbox{side-formul{\ae}}}

\title{Resource-distribution via Boolean constraints}

\author{\begin{tabular}{cc}
James Harland\thanks{Department of Computer Science, Royal Melbourne
Institute of Technology, GPO Box 2476V, Melbourne 3001,
Australia. \emph{Email:} {jah@cs.rmit.edu.au};
\emph{WWW:} {www.cs.rmit.edu.au/$\sim$jah}} &
David Pym\thanks{Department of Computer Science, Queen Mary,
University of London, London E1 4NS, U.K.
\emph{Email:} {pym@dcs.qmw.ac.uk}; \emph{WWW:} {www.dcs.qmw.ac.uk/$\sim$pym}}
    \end{tabular}}

\begin{document}

    \maketitle

\begin{abstract}
We consider the problem of searching for proofs in sequential
presentations of logics with multiplicative (or intensional)
connectives. Specifically, we start with the multiplicative
fragment of linear logic and extend, on the one hand, to linear
logic with its additives and, on the other, to the additives
of the logic of bunched implications, \BI. We give an algebraic
method for calculating the distribution of the side-formul{\ae} in
multiplicative rules which allows the occurrence or non-occurrence
of a formula on a branch of a proof to be determined once sufficient
information is available. Each formula in the conclusion of such a
rule is assigned a Boolean expression. As a search proceeds, a set
of Boolean constraint equations is generated. We show that a
solution to such a set of equations determines a proof corresponding
to the given search. We explain a range of strategies, from the lazy to
the eager, for solving sets of constraint equations. We indicate how to
apply our methods systematically to large family of relevant systems.
\end{abstract}

\section{Introduction}

Proof-search in logics, such as linear logic (LL) \cite{girard87} or
the logic of bunched implications (BI) \cite{OP99,pym99lics,Pym00Mono},
which have multiplicative connectives requires a mechanism by which the
distribution of formul{\ae}, sometimes viewed as \emph{resources},
between different branches of a proof may be calculated.
Such mechanisms are usually specified by
intricate rules of inference which are used to keep track of the
relevant information \cite{hm94,wh95} and are tailored for one
particular distribution mechanism.

We show how a judicious use of Boolean expressions admits
an explicit statement of the resource distribution
problem (unlike in the standard sequent calculus for such logics)
without making a commitment to a particular strategy for
managing the distribution (unlike the tailored inference systems of
\cite{hm94,wh95}). Thus we pursue an \emph{algebraic} approach to this
particular problem of proof-search, allowing a separation of the
specification of the problem from the method of its solution.

The key step in our approach is to assign a Boolean expression to each
of the side-formul{\ae} in multiplicative rules (\ie\ just those
formul{\ae} which require distribution). Constraints on the
possible values of this expression are then generated during the
search process and propagated up the search tree, resulting in
a set of Boolean equations. A successful proof-search
will generate a solution to this set of equations and hence the
corresponding distribution of formul{\ae} across the multiplicative
branches of the proof may be calculated.

We show how a range of different strategies for solving the Boolean
constraints may be considered in this framework, whilst maintaining
the same set of inference rules. Hence the algebraic approach allows us to
separate the inference process from the precise distribution of
formul{\ae} across the branches.

To illustrate the process, consider the $\otimes{\rm R}$ and
$\por{\rm L}$ rules of linear logic \cite{girard87} below.
$$
\infer[\otimes{\rm R}\qquad{\rm and}\qquad]{\Gamma_1 , \Gamma_2 \vdash
\phi_1 \otimes \phi_2 , \Delta_1 , \Delta_2}{\Gamma_1 \vdash \phi_1 , \Delta_1 \quad
                   \Gamma_2 \vdash \phi_2 , \Delta_2}
\infer[\por{\rm L},]{\Gamma_1 , \Gamma_2 , \phi_1 \por \phi_2 \vdash
\Delta_1 , \Delta_2}{\Gamma_1 , \phi_1 \vdash \Delta_1 \quad
                   \Gamma_2 , \phi_2 \vdash \Delta_2}
$$

When considered as rules of deductive inference (\ie\ proofs are
generated from premisses to conclusion), there is no distribution
problem; the multisets of formul{\ae} $\Gamma_1$, $\Gamma_2$, $\Delta_1$
and $\Delta_2$ are known in advance. However, when considered as rules of search
(\ie\ reductive inference), then these four multisets are unknown and
must be calculated in the course of proof-search.

In \S\ref{sec:MLL}, we address this problem for \emph{multiplicative
linear logic}, {\bf MLL}. In particular, we show how the problem of
distributing the \side\ may  be addressed by attaching a Boolean
expression to each such formula together with some simple rules for
the generation of constraints on the value of this expression. The
problem of allocation of the formula to a branch is thus reduced to
determining (in an appropriate manner) a solution to a set of
equations over Boolean expressions. We show that this method is sound
and complete for this logic.

In \S\ref{sec:MALL-EXP}, we extend the results of the previous
section to include the additive and exponential connectives of linear
logic, \ie\ to {\em propositional linear logic}, {\bf PLL}.
This is quite straightforward.

In \S\ref{sec:BI}, we extend the approach of \S\ref{sec:MLL} to the
(propositional fragment of) the \emph{logic of bunched implications},
\BI\ \cite{pym99lics}. This may  be seen as a different extension of {\bf MLL},
in that the additive connectives are introduced in a more intricate
manner. Accordingly the results are correspondingly more subtle in
this case.

In \S\ref{sec:strategies}, we discuss the manner in which the
equations generated by the inference rules may  be systematically
collected and solved. There are three main classes of strategy here:
\emph{lazy}, \emph{eager} and \emph{intermediate}. Lazy strategies are
those in which one branch of the proof-tree is selected and then
followed to completion;the corresponding equations are solved and
then the solution which has been found is broadcast to all other branches. Another
branch is then selected, and the process continues. Such strategies
have properties similar to \emph{depth-first search} and are typically
used in the implementation of linear logic programming languages such
as Lygon \cite{hpw96,wh95,ph94} and Lolli \cite{hm94}. Eager strategies are
those in which all branches are searched in parallel and one single
set of equations is collected. This single set of equations is then
solved once all equations have been found.  Intermediate strategies
solve a fixed number of branches' worth of equations at one time. We
discuss how each of these classes of strategies may be utilized in our
framework.

In \S~\ref{sec:general}, we conclude with an account of the application
of our methods to the family of relevant logics, as described
systematically by Read \cite{Read88}.

We consider our approach to be an algebraic view of proof-search;
in particular, this approach may be considered as a form of
\emph{labelled deduction} \cite{Gabbay95}. Our approach is
similar in spirit to that of \emph{proof-nets} \cite{girard87}
and has its origins in the notion of \emph{path} introduced in
\cite{ph94}; however, it should be noted that in our case we
recover multiplicative consequences from additive rules by the use of
Boolean constraints, rather than recover additive consequences from an
underlying multiplicative system. We have suggested that it is helpful
to consider the formul{\ae} to be distributed as resources and,
indeed, in the case of linear logic there is a very simple sense in
which this is possible: LL can be considered to count the number of
occurrences of a formula. The treatment of resources provided by \BI,
on the other hand, is more delicate. This point is discussed further
in \S~\ref{sec:BI}.

Some previous results on linear logic were presented in \cite{hp97};
in the present paper we develop these results further and extend them
to \BI.

\section{Multiplicatives} \label{sec:MLL}

We begin by restricting our attention to \emph{multiplicative linear
logic} {\bf MLL}; this is a fragment common to both linear logic and
\BI. In addition, throughout the paper we consider only propositional
fragments. Quantifiers pose no problem \emph{per se} and, especially in the
case of linear logic, may be incorporated into what follows in a
straightforward manner.\footnote{Note that quantifiers for linear
logic were included in the earlier paper \cite{hp97}.}
However, the essential issues are more simply described in their
absence.

\begin{define}
The formul{\ae} of (propositional) \emph{multiplicative linear logic} {\bf
MLL} are as follows:

\begin{center}
$p \mid \phi \otimes \phi \mid \phi \por \phi \mid \phi \limp \phi \mid \phi^\perp \mid {\bf 1} \mid \; \perp$
\end{center}
where $p$ is an atom.
\end{define}

Note that we do not include the exponentials here; further discussion
on this point may  be found in \S\ref{sec:MALL-EXP}.

In order to represent the appropriate constraints on the distribution
of formul{\ae}, we do not require arbitrary Boolean expressions. In
particular, we require only those given by the following grammar:

$$e ::= x \,\mid\, \overline{x} \,\mid\, x.e \,\mid\,
\overline{x}.e\;$$
where $x$ is a Boolean variable.

However, the results of this paper are not dependent on the precise
class of formul{\ae}, and hence any class of Boolean expressions which
includes those above will suffice.\footnote{Naturally such
considerations are vital when the complexity of determining the
satisfiability of such expressions is concerned. However, this is a
topic beyond the scope of the present paper.}

\begin{define}
An \emph{annotated} formula is a formula $\phi$ together with a Boolean
expression $e$, denoted as $\phi[e]$.
We denote by $exp(\phi)$ the Boolean expression associated with the
annotated formula $\phi$.
A sequent consisting
entirely of annotated formul{\ae} is known as a \emph{resource sequent}.
\end{define}

In general, the state of the knowledge of the distribution of the
formul{\ae} is characterized by the state of knowledge of the Boolean
variables, with the distribution of the formula known iff the
corresponding Boolean expression has been assigned a value.  Hence in
addition to the proof tree, we maintain an assignment of the Boolean
variables which appear in the proof tree.

\begin{define}

Given a multiset of annotated formul{\ae} $\Delta = \{\phi_1[e_1], \ldots
\phi_n[e_n]\}$ and a total assignment $I$ of the Boolean variables in $\Delta$,
we define $\Delta[I] = \{\phi_1[v_1], \ldots \phi_n[v_n]\}$, where $e_i$ has
the value $v_i$ under $I$.
We denote by $\Delta[I]^1$ the multiset of annotated formul{\ae} $\phi[e]$
in $\Delta[I]$ such that $e$ evaluates to $1$ under $I$.

\end{define}

We will often identify an unannotated formula $\phi$ with the annotated
formula $\phi[1]$ (for instance, in Proposition \ref{prop:mllsound});
it will always be possible to disambiguate such annotations from the context.

\begin{define}
Let $V = \{x_1, x_2, \ldots x_n\}$ be a set of Boolean variables. Then
we denote by $\ov{V}$ the set of Boolean expressions
$\{\ov{x_1}, \ov{x_2}, \ldots \ov{x_n}\}$.
We denote by $\{e\}^n$ the multiset which contains $n$ copies of the
Boolean expression $e$.
\end{define}

\begin{define}\hfill
Let $\Gamma = \{\phi_1[e_1], \phi_2[e_2], \ldots \phi_n[e_n]\}$ be a multiset of
annotated formul{\ae} and let $\{x_1, x_2, \ldots , x_n\}$ be a set of
Boolean variables not occurring in $\Gamma$.
Then we define
$$
\Gamma . \{x_1, x_2, \ldots , x_n\} = \{\phi_1[e_1.x_1],
\phi_2[e_2.x_2], \ldots , \phi_n[e_n.x_n]\}.
$$
\end{define}

We are now in a position to define the inference rules. Roughly
speaking, the Boolean constraints are introduced by the rules which
require multiplicative distribution of formul{\ae} ($\otimes$R, $\por$L,
$\limp$L), are resolved by rules with no premisses (Axiom, $\perp$L,
{\bf 1}R) and are maintained by all the others. Note also that the
principal formula of each rule must be assigned the value 1; hence
the selection of the principal formula also makes some contribution to
the solution of the equations.

\begin{define} \label{def:rules}
We define the following \emph{sequent calculus with constraints} for
{\bf MLL}:

{\small
\begin{tabular}{llll}
& \\
\mbox{\rm Axiom} & $\ds\frac{{\small e_1 = e_2 = 1 \; \; \forall e_3 \in {\rm exp}(\Gamma
\cup \Delta) \; e_3 = 0}} {\Gamma, p[e_1] \vdash p[e_2], \Delta}$ & &
\\
& & & \\

$\bot{\rm L}$ & $\ds\frac{{\small \forall e_2 \in {\rm exp}(\Gamma
\cup\Delta) \; e_2 = 0}}{\Gamma, \bot[e_1] \vdash \Delta}$ $e_1 = 1$ &
$\bot{\rm R}$ & $\ds\frac{\Gamma \vdash \Delta}{\Gamma \vdash \bot[e],
\Delta}$ $e = 1$ \\ & & & \\

$1{\rm L}$ & $\ds\frac{\Gamma \vdash \Delta}{\Gamma, {\bf 1}[e] \vdash
\Delta}$ $e = 1$
&
${\bf 1}{\rm R}$ & $\ds\frac{{\small \forall e_2 \in {\rm exp}(\Gamma \cup
\Delta) \; e_2 = 0}}{\Gamma \vdash {\bf 1}[e_1], \Delta}$ $e_1 = 1$\\ & & & \\

$\por{\rm L}$ & $\ds\frac{\Gamma . V, \phi_1[e] \vdash \Delta . W \quad \Gamma . \ov{V},
\phi_2[e] \vdash \Delta . \ov{W} }{\Gamma, (\phi_1 \por \phi_2)[e] \vdash
\Delta}$ $e = 1$
&
$\por{\rm R}$ & $\ds\frac{\Gamma \vdash \phi_1[e], {\phi_2}[e], \Delta} {\Gamma \vdash (\phi_1
\por \phi_2)[e], \Delta}$ $e = 1$\\ & & & \\

$\otimes{\rm L}$ & $\ds\frac{\Gamma, \phi_1[e], {\phi_2}[e] \vdash
\Delta}{\Gamma, (\phi_1 \otimes \phi_2)[e] \vdash \Delta}$ $e = 1$
&
$\otimes{\rm R}$ & $\ds\frac{\Gamma . V \vdash \phi_1[e], \Delta . W \quad \Gamma . \ov{V}
\vdash \phi_2[e], \Delta . \ov{W} }{\Gamma \vdash (\phi_1 \otimes \phi_2)[e],
\Delta}$ $e = 1$ \\ & & & \\

$\limp{\rm L}$ & $\ds\frac{\Gamma . V \vdash \phi_1[e], \Delta . W \hspace{0.2in} \Gamma
. \ov{V}, \phi_2[e] \vdash \Delta . \ov{W} } {\Gamma, (\phi_1 \limp \phi_2)[e]
\vdash \Delta}$ $e = 1$
&
$\limp{\rm R}$ & $\ds\frac{\Gamma, {\phi_1}[e] \vdash {\phi_2}[e], \Delta} {\Gamma \vdash (\phi_1
\limp \phi_2)[e], \Delta}$ $e = 1$ \\ & & & \\

${\rm L}^\perp$ & $\ds\frac{\Gamma \vdash \phi[e], \Delta}{\Gamma, \phi^\perp[e] \vdash
\Delta}$ $e = 1$
&
${\rm R}^\perp$ & $\ds\frac{\Gamma, \phi[e] \vdash  \Delta}{\Gamma \vdash
\phi^\perp[e], \Delta}$ $e = 1$ \\ & & &\\
\end{tabular}
}

\noindent where the rules $\otimes${\rm R}, $\por${\rm L} and
$\limp${\rm L}
have the side-condition that $V$ and $W$ are disjoint sets of Boolean
variables, none of which occur in $\Gamma$, $\Delta$, $\phi_1$ or $\phi_2$.
We refer to the rules Axiom, {\bf 1}{\rm R} and $\bot${\rm L} as
\emph{leaf} rules.
\end{define}

We then require a proof in the above system to consist of not just the
appropriate proof tree, but also an assignment of the Boolean
variables which occur in it.

\begin{define} \label{def:resource-derivation}
A \emph{resource derivation} is a tree regulated by the
rules of the resource calculus in which each formula of the endsequent
is assigned a distinct Boolean variable, together with a partial assignment
of the Boolean variables appearing in the derivation.

A resource derivation is \emph{total} if its assignment of the
Boolean variables is total. Otherwise, the resource derivation is {\em
partial}. A resource derivation is \emph{closed} if all of the leaves of the
proof tree are instances of one of the leaf rules.
A \emph{resource proof} is a total, closed resource derivation in
which all the Boolean variables in the endsequent and all
principal formul{\ae} are assigned the value $1$.
\end{define}

For notational simplicity, a lack of annotation in any of the rules
of the resource calculus implies that the constraints currently applicable
to the formula are not changed.

Given a resource derivation, it is then straightforward to recover the
corresponding linear proof tree. Note that when the
resource derivation is a resource proof, this linear proof tree will
actually be a proof in the linear sequent calculus.

\begin{define}
Let $R$ be a total resource derivation, with proof tree $T$ and
Boolean assignment $I$.
The \emph{linear proof tree corresponding to R under I}
is the proof tree obtained by deleting from $T$ all formul{\ae}
whose Boolean expression evaluates to $0$ under $I$.
\end{define}

There may  be many resource proofs which have the same
corresponding linear proof --- for example, the linear proof
corresponding to a resource proof of $\Gamma \vdash \Delta$ will be
the same as one corresponding to $\Gamma, \phi[0] \vdash \Delta$ (see
Lemma~\ref{lemma:weak}).

Note that resource proofs do not alter the applicability of the
inference rules. In particular, they \emph{do not} allow us to
construct a ``proof'' of $p \otimes q \vdash p \otimes q$ in which the
right-hand tensor is reduced first: such a proof is prohibited by the
requirement of the $\otimes$L rule that each of $p$, $q$ and $p
\otimes q$ be assigned the same value.

To see this, consider the attempt below at such a resource proof:

$$
\infer[\otimes {\rm R}]{(p \otimes q)[x] \vdash (p \otimes q)[y]}
{
        \infer[\otimes {\rm L}]{(p \otimes q)[x] \vdash p[y]}
                {p[x.z], q[x.z] \vdash p[y]}
        &
        \infer[\otimes {\rm L}]{(p \otimes q)[x.\overline{z}] \vdash q[y]}
                {p[x.\overline{z}], q[x.\overline{z}] \vdash q[y]}
}
$$

Clearly we require that $x=1$ and $y=1$ in order for the rules to be
applied, but but it is impossible to make either leaf of this tree
into an instance of the Axiom rule, due to the impossibility of
finding an appropriate value for $z$.

Resource-proofs are sound and complete with respect to {\bf MLL} proofs.
As usual, soundness reduces to showing that our global
conditions are strong enough to recover proofs from the locally
unsound system.

\begin{prop}[soundness of resource proofs]\label{prop:mllsound}
Let $\Gamma \vdash \Delta$ be a resource sequent in {\bf MLL}.
If $\Gamma \vdash \Delta$ has a resource proof $R$ with Boolean
assignment $I$, then the linear
proof tree corresponding to $R$ is a linear proof of
$\Gamma[I]^1 \vdash \Delta[I]^1$.
\end{prop}

\begin{proof}

By induction on the structure of resource proofs.
Assume that $\Gamma \vdash \Delta$ has a resource proof.

In the base case, the sequent is the conclusion of either the Axiom,
$\perp$L or {\bf 1}R rules.

\begin{itemize}
\item[Axiom:]
We have
$\Gamma, p[e_1] \vdash p[e_2], \Delta$ where $e_1 = e_2 = 1$ and
$\forall e_3 \in {\rm exp}(\Gamma \cup \Delta) e_3 = 0$. Let this
assignment of Boolean variables be $I$. Then
$\Gamma[I]^1, (p[e_1])[I]^1 \vdash (p[e_2])[I]^1, \Delta[I]^1$ is just
$p \vdash p$, and clearly the linear proof tree corresponding to $R$
is a proof of this sequent.

\item[$\perp$L:]
We have
$\Gamma, \perp[e_1] \vdash \Delta$ where $e_1 = 1$ and
$\forall e_2 \in {\rm exp}(\Gamma \cup \Delta) e_2 = 0$. Let this
assignment of Boolean variables be $I$. Then
$\Gamma[I]^1, (\perp[e_1])[I]^1 \vdash \Delta[I]^1$ is just
$\perp \; \vdash $, and clearly the linear proof tree corresponding to $R$
is a proof of this sequent.

\item[{\bf 1}R:]
We have
$\Gamma \vdash {\bf 1}[e_1], \Delta$ where $e_1 = 1$ and
$\forall e_2 \in {\rm exp}(\Gamma \cup \Delta) e_2 = 0$. Let this
assignment of Boolean variables be $I$. Then
$\Gamma[I]^1 \vdash ({\bf 1}[e_1])[I]^1, \Delta[I]^1$ is just
$ \vdash {\bf 1}$, and clearly the linear proof tree corresponding to $R$
is a proof of this sequent.

\end{itemize}

Hence we assume that the result holds for all provable sequents whose
resource proof is no more than a given size.

Consider the last rule used in the proof. There are 10 cases:

\begin{itemize}

\item[{\bf 1}L:]
In this case, we have that
$\Gamma = \Gamma', {\bf 1}[e]$ where $e = 1$, and there is a resource proof of
$\Gamma' \vdash \Delta$. By the hypothesis, there is a linear proof of
$\Gamma'[I]^1 \vdash \Delta[I]^1$ for some Boolean assignment $I$, and
hence there is a linear proof of
$\Gamma'[I]^1, {\bf 1}[1] \vdash \Delta[I]^1$, which is just
$\Gamma[I]^1 \vdash \Delta[I]^1$, as required.

\item[$\perp$R:]
In this case, we have that
$\Delta = \perp[e], \Delta'$ where $e = 1$, and there is a resource proof of
$\Gamma \vdash \Delta'$. By the hypothesis, there is a linear proof of
$\Gamma[I]^1 \vdash \Delta'[I]^1$ for some Boolean assignment $I$, and
hence there is a linear proof of
$\Gamma[I]^1 \vdash \perp[1], \Delta'[I]^1$, which is just
$\Gamma[I]^1 \vdash \Delta[I]^1$, as required.

\item[$\por$L:]
In this case, we have that
$\Gamma = \Gamma', (\phi_1 \por \phi_2)[e]$ where $e = 1$, and
$\Gamma' . V, \phi_1 \vdash \Delta . W$ and
$\Gamma' . \ov{V}, \phi_2 \vdash \Delta . \ov{W}$
both have resource proofs for some disjoint
sets of Boolean variables $V$ and $W$. By the
hypothesis, we have that there are linear proofs of
$(\Gamma' . V)[I]^1, (\phi_1)[I]^1 \vdash (\Delta . W)[I]^1$ and
$(\Gamma' . \ov{V})[I]^1, (\phi_2)[I]^1 \vdash (\Delta . \ov{W})[I]^1$
(recall that $I$ must be an assignment of \emph{all} Boolean variables in the proof), and
so there is a linear proof of
$(\Gamma' . V)[I]^1, (\Gamma; . \ov{V})[I]^1 (\phi_1 \por \phi_2)[I]^1 \vdash
(\Delta . W)[I]^1, (\Delta . \ov{W})[I]^1$
which is just
$\Gamma[I]^1 \vdash \Delta [I]^1$, as required.

\item[$\por$R:]
In this case, we have that
$\Delta = (\phi_1 \por \phi_2)[e],\Delta'$ where $e = 1$, and there is a resource proof of
$\Gamma \vdash \phi_1, \phi_2, \Delta'$. By the hypothesis, there is a linear proof of
$\Gamma[I]^1 \vdash \phi_1[I]^1, \phi_2[I]^1, \Delta'[I]^1$ for some Boolean assignment $I$, and
hence there is a linear proof of
$\Gamma[I]^1 \vdash (\phi_1 \por \phi_2)[I]^1, \Delta'[I]^1$, which is just
$\Gamma[I]^1 \vdash \Delta[I]^1$, as required.

\item[$\otimes$L:]
In this case, we have that
$\Gamma = (\phi_1 \otimes \phi_2)[e],\Gamma'$ where $e = 1$, and there is a resource proof of
$\Gamma', \phi_1, \phi_2 \vdash \Delta$. By the hypothesis, there is a linear proof of
$\Gamma'[I]^1, \phi_1[I]^1, \phi_2[I]^1 \vdash \Delta[I]^1$ for some Boolean assignment $I$, and
hence there is a linear proof of
$\Gamma'[I]^1 (\phi_1 \otimes \phi_2)[I]^1 \vdash \Delta[I]^1$, which is just
$\Gamma'[I]^1 \vdash \Delta[I]^1$, as required.

\item[$\otimes$R:]
In this case, we have that
$\Delta = (\phi_1 \otimes \phi_2)[e], \Delta'$ where $e = 1$, and
$\Gamma . V \vdash \phi_1, \Delta' . W$ and
$\Gamma . \ov{V} \vdash \phi_2, \Delta' . \ov{W}$
both have resource proofs for some disjoint
sets of Boolean variables $V$ and $W$. By the
hypothesis, we have that there are linear proofs of
$(\Gamma . V)[I]^1 \vdash (\phi_1)[I]^1, (\Delta' . W)[I]^1$ and
$(\Gamma . \ov{V})[I]^1 \vdash (\phi_2)[I]^1, (\Delta' . \ov{W})[I]^1$
(recall that $I$ must be an assignment of \emph{all} Boolean variables in the proof), and
so there is a linear proof of
$(\Gamma . V)[I]^1, (\Gamma . \ov{V})[I]^1 \vdash (\phi_1 \otimes
\phi_2)[I]^1, (\Delta' . W)[I]^1, (\Delta' . \ov{W})[I]^1$
which is just
$\Gamma[I]^1 \vdash \Delta [I]^1$, as required.

\item[$\limp$L:]
In this case, we have that
$\Gamma = (\phi_1 \limp \phi_2)[e1], \Gamma'$ where $e = 1$, and
$\Gamma' . V \vdash \phi_1, \Delta . W$ and
$\Gamma' . \ov{V}, \phi_2 \vdash \Delta . \ov{W}$
both have resource proofs for some disjoint
sets of Boolean variables $V$ and $W$. By the
hypothesis, we have that there are linear proofs of
$(\Gamma' . V)[I]^1 \vdash (\phi_1)[I]^1, (\Delta . W)[I]^1$ and
$(\Gamma' . \ov{V})[I]^1, (\phi_2)[I]^1 \vdash (\Delta . \ov{W})[I]^1$
(recall that $I$ must be an assignment of \emph{all} Boolean variables in the proof), and
so there is a linear proof of
$(\Gamma' . V)[I]^1, (\Gamma' . \ov{V})[I]^1 (\phi_1 \limp \phi_2)[I]^1 \vdash
(\Delta . W)[I]^1, (\Delta . \ov{W})[I]^1$
which is just
$\Gamma[I]^1 \vdash \Delta [I]^1$, as required.

\item[$\limp$R:]
In this case, we have that
$\Delta = (\phi_1 \limp \phi_2)[e],\Delta'$ where $e = 1$, and there is a resource proof of
$\Gamma, \phi_1 \vdash \phi_2, \Delta'$. By the hypothesis, there is a linear proof of
$\Gamma[I]^1, \phi_1[I]^1 \vdash  \phi_2[I]^1, \Delta'[I]^1$ for some Boolean assignment $I$, and
hence there is a linear proof of
$\Gamma[I]^1 \vdash (\phi_1 \limp \phi_2)[I]^1, \Delta'[I]^1$, which is just
$\Gamma[I]^1 \vdash \Delta[I]^1$, as required.

\item[$L^\perp$:]
In this case, we have that
$\Gamma = (\phi^\perp)[e], \Gamma'$ where $e = 1$, and there is a resource proof of
$\Gamma' \vdash \phi, \Delta$. By the hypothesis, there is a linear proof of
$\Gamma'[I]^1 \vdash \phi[I]^1, \Delta[I]^1$ for some Boolean assignment $I$, and
hence there is a linear proof of
$\Gamma'[I]^1, (\phi^\perp)[I]^1 \vdash \Delta[I]^1$, which is just
$\Gamma[I]^1 \vdash \Delta[I]^1$, as required.

\item[$R^\perp$:]
In this case, we have that
$\Delta = (\phi^\perp)[e], \Delta'$ where $e = 1$, and there is a resource proof of
$\Gamma, \phi \vdash \Delta'$. By the hypothesis, there is a linear proof of
$\Gamma[I]^1, \phi[I]^1 \vdash \Delta'[I]^1$ for some Boolean assignment $I$, and
hence there is a linear proof of
$\Gamma[I]^1 \vdash (\phi^\perp)[I]^1, \Delta'[I]^1$, which is just
$\Gamma[I]^1 \vdash \Delta[I]^1$, as required.

\end{itemize}

\end{proof}

In order to show the completeness of resource proofs, we require the
following simple lemma:

\begin{lemma} \label{lemma:weak}
Let $\Gamma \vdash \Delta$ be a resource sequent in {\bf MLL}.
If $\Gamma \vdash \Delta$ has a closed resource derivation $R$, then
$\Gamma, \phi[0] \vdash \Delta$ and $\Gamma \vdash \phi[0], \Delta$
also have closed resource derivations $R_1$ and $R_2$ respectively.
Moreover, for any total assignment $I$ of the Boolean variables in
$R$, the linear proof tree corresponding to $R$ under $I$ is the same
as the linear proof trees corresponding to $R_1$ and $R_2$
respectively under $I$.
\end{lemma}

\begin{proof}
By induction on the structure of resource derivations.
Assume that $\Gamma \vdash \Delta$ has a closed resource derivation.

In the base case, the sequent is the conclusion of either the Axiom,
$\perp$L or {\bf 1}R rules, and it is clear that the addition of
$\phi[0]$ to either the antecedent or the succedent of $\Gamma \vdash
\Delta$ will result in a closed resource derivation of the appropriate
sequent.  Clearly the linear proof tree property is satisfied in these
cases.

Hence we assume that the result holds for all provable sequents whose
resource proof is no more than a given size.

Consider the last rule used in the proof. There are 10 cases, of which
we only give the argument for $\otimes$L and $\otimes$R; the others
are similar.

\begin{itemize}

\item[$\otimes$L:]

In this case, we have that
$\Gamma = (\phi_1 \otimes \phi_2)[e], \Gamma'$ where $e = 1$, and there is a closed resource derivation of
$\Gamma', \phi_1[1], \phi_2[1] \vdash \Delta$. By the hypothesis, there are closed
resource derivations of both
$\Gamma', \phi_1[1], \phi_2[1], \phi[0] \vdash \Delta$  and
$\Gamma', \phi_1[1], \phi_2[1] \vdash \Delta, \phi[0]$, and so there are closed
resource derivations of
$\Gamma', (\phi_1 \otimes \phi_2)[1], \phi[0] \vdash \Delta$ and
$\Gamma', (\phi_1 \otimes \phi_2)[1] \vdash \Delta, \phi[0]$, as required.

Clearly the linear proof tree property is satisfied in this case.

\item[$\otimes$R:]

In this case, we have that
$\Delta = (\phi_1 \otimes \phi_2)[e], \Delta'$ where $e = 1$, and
$\Gamma . V \vdash \phi_1[1], \Delta' . W$ and
$\Gamma . \ov{V} \vdash \phi_2[1], \Delta' . \ov{W}$
both have closed resource derivations for some disjoint
sets of Boolean variables $V$ and $W$. By the
hypothesis, there are closed resource derivations of both
$\Gamma . V, \phi[0] \vdash \phi_1[1], \Delta' . W$ and
$\Gamma . \ov{V}, \phi[0] \vdash \phi_2[1], \Delta' . \ov{W}$
and of both
$\Gamma . V \vdash \phi_1[1], \phi[0], \Delta' . W$ and
$\Gamma . \ov{V} \vdash \phi_2[1], \phi[0], \Delta' . \ov{W}$.
Hence there are disjoint sets of variables $V'$ and $W'$ such that $V
\subset V'$ and $W \subset W'$ such that
$(\Gamma \cup \phi[0]) . V' \vdash \phi_1[1], \Delta' . W$ and
$(\Gamma \cup \phi[0]). \ov{V'} \vdash \phi_2[1], \Delta' . \ov{W}$
both have closed resource derivations, as do
$\Gamma . V \vdash \phi_1[1], (\Delta' \cup \phi[0]) . W'$ and
$\Gamma . \ov{V} \vdash \phi_2[1], (\Delta' \cup \phi[0]) . \ov{W'}$.
Thus we have closed resource derivations of both
$\Gamma, \phi[0] \vdash (\phi_1 \otimes \phi_2)[1], \Delta'$ and
$\Gamma \vdash (\phi_1 \otimes \phi_2)[1], \Delta', \phi[0]$, as required.

Clearly the linear proof tree property is satisfied in this case.

\end{itemize}

\end{proof}

We are now in a position to show the completeness of resource proofs
for {\bf MLL}.

\begin{prop}[completeness of resource proofs]\label{prop:mllcomplete}
Let $\Gamma \vdash \Delta$ be a sequent in {\bf MLL}.
If $\Gamma \vdash \Delta$ has a proof $\Phi$ in {\bf MLL}, then there
are disjoint sets of Boolean variables $V$ and $W$ such that
$\Gamma . V \vdash \Delta . W$ has a resource proof $R$ and the linear proof
tree corresponding to $R$ is $\Phi$.
\end{prop}

\begin{proof}

By induction on the structure of resource proofs.
Assume that $\Gamma \vdash \Delta$ has a resource proof.

In the base case, the sequent is the conclusion of either the Axiom,
$\perp$L or {\bf 1}R rules.

\begin{itemize}
\item[Axiom:]
We have $p \vdash p$, so clearly there is a
resource proof of $p[x] \vdash p[y]$ and the linear proof
corresponding to this resource proof is $\Phi$.

\item[$\perp$L:]
We have $\perp \; \vdash $,  so clearly there is a
resource proof of $\perp[x]\vdash $ and the linear proof
corresponding to this resource proof is $\Phi$.

\item[{\bf 1}R:]
We have $\vdash {\bf 1}$, so clearly there is a
resource proof of $\vdash {\bf 1}[x]$ and the linear proof
corresponding to this resource proof is $\Phi$.

\end{itemize}

Hence we assume that the result holds for all provable sequents whose
resource proof is no more than a given size.

Consider the last rule used in the proof. There are ten cases.

\begin{itemize}

\item[{\bf 1}L:]
In this case, we have that
$\Gamma = \Gamma', {\bf 1}$, and there is a proof of
$\Gamma' \vdash \Delta$ which is a subproof of $\Phi$.
By the hypothesis, there are disjoint sets of
variables $V$ and $W$ such that there is a resource proof of
$\Gamma'.V \vdash \Delta.W$, and
hence there is a resource proof of
$\Gamma'.V, {\bf 1}[1] \vdash \Delta.W$, and
clearly the linear proof corresponding to this resource proof is $\Phi$.

\item[$\perp$R:]
In this case, we have that
$\Delta = \Delta', \perp$, and there is a proof of
$\Gamma \vdash \Delta'$ which is a subproof of $\Phi$. By the hypothesis, there are disjoint sets of
variables $V$ and $W$ such that there is a resource proof of
$\Gamma.V \vdash \Delta'.W$, and
hence there is a resource proof of
$\Gamma'.V \vdash \perp[1], \Delta.W$, and
clearly the linear proof corresponding to this resource proof is $\Phi$.

\item[$\por$L:]
In this case,  we have that
$\Gamma = \phi_1 \por \phi_2, \Gamma_1 \Gamma_2$, $\Delta = \Delta_1, \Delta_2$ such that
$\Gamma_1, \phi_1 \vdash \Delta_1$ and
$\Gamma_2, \phi_2 \vdash \Delta_2$
both have proofs in the linear sequent
calculus which are subproofs of $\Phi$. Hence by the hypothesis there are disjoint sets of Boolean
variables $V_1, V_2, W_1, W_2$ such that
$\Gamma_1 . V_1, \phi_1[1] \vdash \Delta_1 . W_1$ and
$\Gamma_2 . V_2, \phi_2[1] \vdash \Delta_2 . W_2$ have resource proofs,
(and moreover the linear proofs corresponding to each resource proof is the appropriate subproof of $\Phi$)
and so by Lemma \ref{lemma:weak}, there are closed resource derivations of
$\Gamma_1 . V_1, \Gamma_2 . \{0\}^n, \phi_1[1] \vdash \Delta_1 . W_1, \Delta_2 . \{0\}^n$ and
$\Gamma_1 . \{0\}^n, \Gamma_2 . V_2, \phi_2[1] \vdash \Delta_1 . \{0\}^n, \Delta_2 . W_2$.
Hence there are new disjoint sets of Boolean variables (\ie\ not
occurring anywhere in the above two resource sequents) $V$ and $W$
and a total assignment $I$ of $V \cup W$ such that
$(\Gamma_1 . V_1, \Gamma_2 . V_2). V, \phi_1[1] \vdash (\Delta_1 . W_1,
\Delta_2 . W_2). W$ and
$(\Gamma_1 . V_1, \Gamma_2 . V_2). \ov{V}, \phi_2[1] \vdash (\Delta_1
. W_1,\Delta_2 . W_2) . \ov{W}$
have resource proofs, and so there is a resource proof of
$\Gamma_1 . V_1, \Gamma_2 . V_2, (\phi_1 \por \phi_2)[1] \vdash \Delta_1
. W_1, \Delta_2 . W_2$, \ie\ $\Gamma . V', (\phi_1 \por \phi_2)[1] \vdash
\Delta . W'$
for some disjoint sets of  Boolean variables $V'$ and $W'$, and
clearly the linear proof corresponding to this resource proof is $\Phi$.

\item[$\por$R:]
In this case, we have that
$\Delta = \phi_1 \por \phi_2,\Delta'$, and there is a proof of
$\Gamma \vdash \phi_1, \phi_2, \Delta'$ which is a subproof of $\Phi$. By the hypothesis, there are disjoint sets of
variables $V$ and $W$ such that there is a resource proof of
$\Gamma.V \vdash \phi_1[1], \phi_2[1], \Delta'.W$, and
hence there is a resource proof of
$\Gamma'.V \vdash (\phi_1 \por \phi_2)[1], \Delta.W$, and
clearly the linear proof corresponding to this resource-proof is $\Phi$.

\item[$\otimes$L:]
In this case, we have that
$\Gamma = \phi_1 \otimes \phi_2, \Gamma'$, and there is a proof of
$\Gamma', \phi_1, \phi_2 \vdash \Delta$ which is a subproof of $\Phi$. By the hypothesis, there are disjoint sets of
variables $V$ and $W$ such that there is a resource proof of
$\Gamma'.V, \phi_1[1], \phi_2[1] \vdash \Delta.W$, and
hence there is a resource proof of
$\Gamma'.V, (\phi_1 \otimes \phi_2)[1] \vdash  \Delta.W$, and
clearly the linear proof corresponding to this resource proof is $\Phi$.

\item[$\otimes$R:]
In this case,  we have that
$\Gamma = \Gamma_1, \Gamma_2$, $\Delta = \phi_1 \otimes \phi_2, \Delta_1, \Delta_2$ such that
$\Gamma_1 \vdash \phi_1, \Delta_1$ and
$\Gamma_2 \vdash \phi_2, \Delta_2$
both have proofs in the linear sequent
calculus which are subproofs of $\Phi$. Hence by the hypothesis there are disjoint sets of Boolean
variables $V_1, V_2, W_1, W_2$ such that
$\Gamma_1 . V_1 \vdash \phi_1[1], \Delta_1 . W_1$ and
$\Gamma_2 . V_2 \vdash \phi_2[1], \Delta_2 . W_2$ have resource proofs,
(and moreover the linear proofs corresponding to each resource proof is the appropriate subproof of $\Phi$)
and so by Lemma \ref{lemma:weak}, there are closed
resource derivations of
$\Gamma_1 . V_1, \Gamma_2 . \{0\}^n \vdash \phi_1[1], \Delta_1 . W_1, \Delta_2 . \{0\}^n$ and
$\Gamma_1 . \{0\}^n, \Gamma_2 . V_2 \vdash \phi_2[1], \Delta_1 . \{0\}^n, \Delta_2 . W_2$.
Hence there are new disjoint sets of Boolean variables (\ie\ not
occurring anywhere in the above two resource sequents) $V$ and $W$
and a total assignment $I$ of $V \cup W$ such that
$(\Gamma_1 . V_1, \Gamma_2 . V_2). V \vdash \phi_1[1], (\Delta_1 . W_1,  \Delta_2 . W_2). W$ and
$(\Gamma_1 . V_1, \Gamma_2 . V_2). \ov{V} \vdash \phi_2[1], (\Delta_1
. W_1,\Delta_2 . W_2) . \ov{W}$
have resource proofs, and so there is a resource proof of
$\Gamma_1 . V_1, \Gamma_2 . V_2 \vdash (\phi_1 \otimes \phi_2)[1], \Delta_1
. W_1, \Delta_2 . W_2$, \ie\ $\Gamma . V' \vdash (\phi_1 \otimes \phi_2)[1], \Delta . W'$
for some disjoint sets of  Boolean variables $V'$ and $W'$, and clearly the
linear proof corresponding to this resource proof is $\Phi$.

\item[$\limp$L:]
In this case,  we have that
$\Gamma = \phi_1 \limp \phi_2, \Gamma_1, \Gamma_2$, $\Delta = \Delta_1, \Delta_2$ such that
$\Gamma_1 \vdash \phi_1, \Delta_1$ and
$\Gamma_2, \phi_2 \vdash \Delta_2$
both have proofs in the linear sequent
calculus which are subproofs of $\Phi$. Hence by the hypothesis there are disjoint sets of Boolean
variables $V_1, V_2, W_1, W_2$ such that
$\Gamma_1 . V_1 \vdash \phi_1[1], \Delta_1 . W_1$ and
$\Gamma_2 . V_2, \phi_2[1] \vdash \Delta_2 . W_2$ have resource proofs,
(and moreover the linear proofs corresponding to each resource proof is the appropriate subproof of $\Phi$)
and so by Lemma \ref{lemma:weak}, there are closed resource derivations of
$\Gamma_1 . V_1, \Gamma_2 . \{0\}^n \vdash \phi_1[1], \Delta_1 . W_1, \Delta_2 . \{0\}^n$ and
$\Gamma_1 . \{0\}^n, \Gamma_2 . V_2, \phi_2[1] \vdash \Delta_1 . \{0\}^n, \Delta_2 . W_2$.
Hence there are new disjoint sets of Boolean variables (\ie\ not
occurring anywhere in the above two resource sequents) $V$ and $W$
and a total assignment $I$ of $V \cup W$ such that
$(\Gamma_1 . V_1, \Gamma_2 . V_2). V \vdash \phi_1[1], (\Delta_1 . W_1,
\Delta_2 . W_2). W$ and
$(\Gamma_1 . V_1, \Gamma_2 . V_2). \ov{V}, \phi_2[1] \vdash (\Delta_1
. W_1,\Delta_2 . W_2) . \ov{W}$
have resource proofs, and so there is a resource proof of
$\Gamma_1 . V_1, \Gamma_2 . V_2, (\phi_1 \limp \phi_2)[1] \vdash \Delta_1
. W_1, \Delta_2 . W_2$, \ie\ $\Gamma . V', (\phi_1 \limp \phi_2)[1] \vdash
\Delta . W'$
for some disjoint sets of  Boolean variables $V'$ and $W'$, and
clearly the linear proof corresponding to this resource proof is $\Phi$.

\item[$\limp$R:]
In this case, we have that
$\Delta = \phi_1 \limp \phi_2,\Delta'$, and there is a proof of
$\Gamma, \phi_1 \vdash \phi_2, \Delta'$ which is a subproof of $\Phi$. By
the hypothesis, there are disjoint sets of
variables $V$ and $W$ such that there is a resource proof of
$\Gamma.V, \phi_1[1] \vdash \phi_2[1], \Delta'.W$, and
hence there is a resource proof of
$\Gamma'.V \vdash (\phi_1 \limp \phi_2)[1], \Delta.W$, and
clearly the linear proof corresponding to this resource proof is $\Phi$.

\item[$L^\perp$:]
In this case, we have that
$\Gamma = \phi^\perp, \Gamma'$, and there is a proof of
$\Gamma' \vdash \phi, \Delta$ which is a subproof of $\Phi$. By the hypothesis,
there are disjoint sets of
variables $V$ and $W$ such that there is a resource proof of
$\Gamma'.V \vdash \phi[1], \Delta.W$, and
hence there is a resource proof of
$\Gamma'.V, (\phi^\perp)[1] \vdash \Delta.W$, and
clearly the linear proof corresponding to this resource proof is $\Phi$.

\item[$R^\perp$:]
In this case, we have that $\Delta = \phi^\perp, \Delta'$, and there is a
proof of $\Gamma, \phi \vdash \Delta'$ which is a subproof of $\Phi$. By
the hypothesis, there are disjoint sets of variables $V$ and $W$ such
that there is a resource proof of $\Gamma.V, \phi[1] \vdash \Delta'.W$, and
hence there is a resource proof of $\Gamma.V \vdash (\phi^\perp)[1], \Delta'.W$,
and clearly the linear proof corresponding to this resource proof is $\Phi$.
\end{itemize}
\end{proof}

The reader will have noted that in the application of any rule, it is
always necessary for the expression for the principal formula of the
rule to be 1. In what follows we will omit an explicit statement of
this requirement when it is obvious from the context.

\section{Linear additives and exponentials} \label{sec:MALL-EXP}

We now consider how to extend the approach of \S2 to include additives
and the exponentials of linear logic. As noted above, we only consider
propositional fragments in this paper; the quantifiers \emph{per se} provide
no great technical problem (indeed, see \cite{hp97} for a basic treatment). 

Hence we now consider the class of formul{\ae} defined in the following
definition:

\begin{define}
The formul{\ae} of \emph{propositional linear logic} {\bf PLL} are as follows:

\begin{center}
$p \mid \phi \otimes \phi \mid \phi \por \phi \mid \phi \limp \phi \mid \phi \with \phi \mid \phi
\oplus \phi \mid !\phi \mid ?\phi \mid \phi^\perp \mid {\bf 1} \mid \; \perp \mid {\bf 0} \mid \top$
\end{center}
where $p$ is an atom.
\end{define}

It is straightforward to extend the rules of resource derivations to
this class of formul{\ae}, as below.

One point to note is that the Boolean expression technique discussed
above may be used to characterize the choice of formula in the
$\oplus$R and $\with$L rules. For example, the $\oplus$R rule requires
a choice to be made between $\phi_1$ and $\phi_2$, as below.

$$
\ds \oplus{\rm R} \; \frac{\Gamma \vdash \phi_i, \Delta}{\Gamma \vdash \phi_1
\oplus \phi_2, \Delta}
$$

However, it is clear that by attaching a Boolean variable to $\phi_1$ and
its complement to $\phi_2$ we are able to use the same constraint
technique as above. This requires the rule to be reformulated as

$$
\ds \oplus{\rm R} \; \frac{\Gamma \vdash \phi_1[x], \phi_2[\ov{x}], \Delta}
{\Gamma \vdash (\phi_1 \oplus \phi_2)[1], \Delta}
$$

Similar remarks apply to the $\with$L rule.

The exponentials are also straightforward, as they do not introduce
any distribution problems. It should be noted, though, that the
technique of attaching Boolean expressions to formul{\ae} may be adapted
to whatever approach is taken to the rules involving exponentials. For
example, it is well-known that the weakening rules (W!L and W?R) may
be permuted upwards, and hence incorporated into the leaf rules.

Also, the rules !L and C!L may be combined into one, as below \cite{Tam94}:

$$
\ds
\frac{\Gamma, \phi, !\phi \vdash \Delta}{\Gamma, !\phi \vdash \Delta}
$$

In either case, it is not hard to see how to adapt the
resource versions of these rules to incorporate these
adaptations. Hence here we consider only the ``naive'' exponential
rules, as variations of them, as mentioned above, constitute only
minor changes to the rules below.

\begin{define} \label{def:rulesPLL}
We define the following \emph{sequent calculus with constraints} for
{\bf PLL}:

{\small
\begin{tabular}{llll}
& \\
Axiom & $\ds\frac{{\small e_1 = e_2 = 1 \; \; \forall e_3 \in {\rm exp}(\Gamma
\cup \Delta) \; e_3 = 0}} {\Gamma, p[e_1] \vdash p[e_2], \Delta}$ & &
\\
& & & \\

$\bot{\rm L}$ & $\ds\frac{{\small \forall e_2 \in {\rm exp}(\Gamma
\cup\Delta) \; e_2 = 0}}{\Gamma, \bot[e_1] \vdash \Delta}$ $e_1 = 1$ &
$\bot{\rm R}$ & $\ds\frac{\Gamma \vdash \Delta}{\Gamma \vdash \bot[e],
\Delta}$ $e = 1$ \\ & & & \\

$1{\rm L}$ & $\ds\frac{\Gamma \vdash \Delta}{\Gamma, {\bf 1}[e] \vdash
\Delta}$ $e = 1$
&
${\bf 1}{\rm R}$ & $\ds\frac{{\small \forall e_2 \in {\rm exp}(\Gamma \cup
\Delta) \; e_2 = 0}}{\Gamma \vdash {\bf 1}[e_1], \Delta}$ $e_1 = 1$\\ & & & \\

${\bf 0}{\rm L}$ & $\ds\frac{}{\Gamma, {\bf 0}[e_1] \vdash \Delta}$
$e_1 = 1$ &
$\top{\rm R}$ & $\ds\frac{{\small e_1 = 1}}
{\Gamma \vdash  \top[e_1] \Delta}$ \\ & & & \\

$\por{\rm L}$ & $\ds\frac{\Gamma . V, \phi_1[e] \vdash \Delta . W \quad \Gamma . \ov{V},
\phi_2[e] \vdash \Delta . \ov{W} }{\Gamma, (\phi_1 \por \phi_2)[e] \vdash
\Delta}$ $e = 1$
&
$\por{\rm R}$ & $\ds\frac{\Gamma \vdash \phi_1[e], {\phi_2}[e], \Delta} {\Gamma \vdash (\phi_1
\por \phi_2)[e], \Delta}$ $e = 1$\\ & & & \\

$\otimes{\rm L}$ & $\ds\frac{\Gamma, \phi_1[e], {\phi_2}[e] \vdash
\Delta}{\Gamma, (\phi_1 \otimes \phi_2)[e] \vdash \Delta}$ $e = 1$
&
$\otimes{\rm R}$ & $\ds\frac{\Gamma . V \vdash \phi_1[e], \Delta . W \quad \Gamma . \ov{V}
\vdash \phi_2[e], \Delta . \ov{W} }{\Gamma \vdash (\phi_1 \otimes \phi_2)[e],
\Delta}$ $e = 1$ \\ & & & \\

$\limp{\rm L}$ & $\ds\frac{\Gamma . V \vdash \phi_1[e], \Delta . W \hspace{0.2in} \Gamma
. \ov{V}, \phi_2[e] \vdash \Delta . \ov{W} } {\Gamma, (\phi_1 \limp \phi_2)[e]
\vdash \Delta}$ $e = 1$
&
$\limp{\rm R}$ & $\ds\frac{\Gamma, {\phi_1}[e] \vdash {\phi_2}[e], \Delta} {\Gamma \vdash (\phi_1
\limp \phi_2)[e], \Delta}$ $e = 1$ \\ & & & \\

${\rm L}^\perp$ & $\ds\frac{\Gamma \vdash \phi[e], \Delta}{\Gamma, \phi^\perp[e] \vdash
\Delta}$ $e = 1$
&
${\rm R}^\perp$ & $\ds\frac{\Gamma, \phi[e] \vdash  \Delta}{\Gamma \vdash
\phi^\perp[e], \Delta}$ $e = 1$ \\ & & &\\

$\oplus{\rm L}$ & $\ds\frac{\Gamma, \phi_1[e] \vdash \Delta \hspace{0.2in} \Gamma, \phi_2[e]
\vdash \Delta}
{\Gamma, (\phi_1 \oplus \phi_2)[e] \vdash \Delta}$ $e = 1$
&
$\oplus{\rm R}$ & $\ds\frac{\Gamma \vdash \phi_1[x], \phi_2[\ov{x}], \Delta} {\Gamma \vdash (\phi_1
\oplus \phi_2)[e], \Delta}$ $e = 1$ \\ & & & \\

$\with{\rm L}$ & $\ds\frac{\Gamma, \phi_1[x], \phi_2[\ov{x}] \vdash \Delta}{\Gamma, (\phi_1 \with \phi_2)[e] \vdash \Delta}$ $e = 1$
&
$\with{\rm R}$ & $\ds\frac{\Gamma \vdash \phi_1[e], \Delta \quad \Gamma \vdash \phi_2[e], \Delta}
{\Gamma \vdash (\phi_1 \with \phi_2)[e], \Delta}$ $e = 1$ \\ & & & \\

$!{\rm L}$ & $\ds\frac{\Gamma, \phi[e] \vdash \Delta}{\Gamma, !\phi[e] \vdash \Delta}$ $e = 1$
&
$!{\rm R}$ & $\ds\frac{!\Gamma \vdash \phi[e], ?\Delta}{!\Gamma \vdash !\phi[e], ?\Delta}$ $e = 1$ \\ & & & \\

$?{\rm L}$ & $\ds\frac{!\Gamma, \phi[e] \vdash ?\Delta}{!\Gamma, ?\phi[e] \vdash ?\Delta}$ $e = 1$
&
$?{\rm R}$ & $\ds\frac{\Gamma \vdash \phi[e], \Delta}{\Gamma \vdash ?\phi[e], \Delta}$ $e = 1$ \\ & & & \\

$W!{\rm L}$ & $\ds\frac{\Gamma \vdash \Delta}{\Gamma, !\phi[e] \vdash \Delta}$ $e = 1$
&
$W?{\rm R}$ & $\ds\frac{\Gamma \vdash \Delta}{\Gamma \vdash ?\phi[e], \Delta}$ $e = 1$ \\ & & & \\

$C!{\rm L}$ & $\ds\frac{\Gamma, !\phi[e], !\phi[e] \vdash \Delta}{\Gamma, !\phi[e] \vdash \Delta}$ $e = 1$
&
$C?{\rm R}$ & $\ds\frac{\Gamma \vdash ?\phi[e], ?\phi[e], \Delta}{\Gamma \vdash ?\phi[e], \Delta}$ $e = 1$ \\ & & & \\

\end{tabular}
}

\noindent where the rules $\otimes$R, $\por$L and $\limp$L
have the side-condition that $V$ and $W$ are disjoint sets of Boolean
variables, none of which occur in $\Gamma$, $\Delta$, $\phi_1$ or $\phi_2$.

\end{define}

It is then straightforward to show the soundness of resource proofs
for {\bf PLL}.

\begin{prop}[soundness of resource proofs]\label{prop:maisound}
Let $\Gamma \vdash \Delta$ be a resource sequent in {\bf PLL}.
If $\Gamma \vdash \Delta$ has a resource proof $R$ with Boolean
assignment $I$, then the linear
proof tree corresponding to $R$ is a linear proof of
$\Gamma[I]^1 \vdash \Delta[I]^1$.
\end{prop}

\begin{proof}

By induction on the structure of resource proofs.
Assume that $\Gamma \vdash \Delta$ has a resource proof.

In the base case, the sequent is the conclusion of either the Axiom,
$\perp$L, {\bf 1}R, {\bf 0}L or $\top$R rules, and it is clear that
the result holds in each of these cases.

Hence we assume that the result holds for all provable sequents whose
resource proof is no more than a given size.

Consider the last rule used in the proof. We only give the argument
for the cases $\otimes$L, $\otimes$R, $\with$L, $\with$R, C!L; the
others are similar.

\begin{itemize}

\item[$\with$L:]
In this case, we have that
$\Gamma = (\phi_1 \with  \phi_2)[1],\Gamma'$, and there is a resource proof of
$\Gamma', \phi_1[x], \phi_2[\ov{x}] \vdash \Delta$. By the hypothesis, there is a
linear proof of
$\Gamma'[I]^1, (\phi_1[x])[I]^1, (\phi_2[\ov{x}])[I]^1 \vdash \Delta[I]^1$
for some Boolean assignment $I$, which is clearly either
$\Gamma'[I]^1, \phi_1 \vdash \Delta[I]^1$ or
$\Gamma'[I]^1, \phi_2 \vdash \Delta[I]^1$,
as required.

\item[$\with$R:]
In this case, we have that
$\Delta = (\phi_1 \with \phi_2)[1], \Delta'$, and
$\Gamma \vdash \phi_1[1], \Delta'$ and
$\Gamma \vdash \phi_2[1], \Delta'$
both have resource proofs.
By the hypothesis, we have that there are linear proofs of
$\Gamma[I]^1 \vdash (\phi_1)[I]^1, \Delta'[I]^1$ and
$\Gamma[I]^1 \vdash (\phi_2)[I]^1, \Delta'[I]^1$,
and so there is a linear proof of
$\Gamma[I]^1 \vdash (\phi_1 \with \phi_2)[I]^1, \Delta'[I]^1$
which is just
$\Gamma[I]^1 \vdash \Delta [I]^1$, as required.

\item[$\otimes$L:]
In this case, we have that
$\Gamma = (\phi_1 \otimes \phi_2)[1],\Gamma'$, and there is a resource proof of
$\Gamma', \phi_1, \phi_2 \vdash \Delta$. By the hypothesis, there is a linear proof of
$\Gamma'[I]^1, \phi_1[I]^1, \phi_2[I]^1 \vdash \Delta[I]^1$ for some Boolean assignment $I$, and
hence there is a linear proof of
$\Gamma'[I]^1 (\phi_1 \otimes \phi_2)[I]^1 \vdash \Delta[I]^1$, which is just
$\Gamma'[I]^1 \vdash \Delta[I]^1$, as required.

\item[$\otimes$R:]
In this case, we have that
$\Delta = (\phi_1 \otimes \phi_2)[1], \Delta'$, and
$\Gamma . V \vdash \phi_1[1], \Delta' . W$ and
$\Gamma . \ov{V} \vdash \phi_2[1], \Delta' . \ov{W}$
both have resource proofs for some disjoint
sets of Boolean variables $V$ and $W$. By the
hypothesis, we have that there are linear proofs of
$(\Gamma . V)[I]^1 \vdash (\phi_1)[I]^1, (\Delta' . W)[I]^1$ and
$(\Gamma . \ov{V})[I]^1 \vdash (\phi_2)[I]^1, (\Delta' . \ov{W})[I]^1$
(recall that $I$ must be an assignment of \emph{all} Boolean variables in the proof), and
so there is a linear proof of
$(\Gamma . V)[I]^1, (\Gamma . \ov{V})[I]^1 \vdash (\phi_1 \otimes
\phi_2)[I]^1, (\Delta' . W)[I]^1, (\Delta' . \ov{W})[I]^1$
which is just
$\Gamma[I]^1 \vdash \Delta [I]^1$, as required.

\item[C!L:]
In this case, we have that
$\Gamma = !\phi[1], \Gamma'$, and there is a resource proof of
$\Gamma', !\phi[1], !\phi[1] \vdash \Delta$. By the hypothesis, there is a
linear proof of
$\Gamma'[I]^1, !\phi[I]^1, !\phi[I]^1 \vdash \Delta[I]^1$ for some Boolean assignment $I$, and
hence there is a linear proof of
$\Gamma'[I]^1, !\phi[I]^1 \vdash \Delta[I]^1$, which is just
$\Gamma[I]^1 \vdash \Delta[I]^1$, as required.
\end{itemize}

\end{proof}

As previously, we require the following lemma in order to show the
completeness of resource proofs for {\bf PLL}.

\begin{lemma} \label{lemma:maiweak}
Let $\Gamma \vdash \Delta$ be a resource sequent in {\bf PLL}.
If $\Gamma \vdash \Delta$ has a closed resource derivation $R$, then
$\Gamma, \phi[0] \vdash \Delta$ and $\Gamma \vdash \phi[0], \Delta$
also have closed resource derivations $R_1$ and $R_2$ respectively.
Moreover, for any total assignment $I$ of the Boolean variables in
$R$, the linear proof tree corresponding to $R$ under $I$ is the same
as the linear proof trees corresponding to $R_1$ and $R_2$
respectively under $I$.
\end{lemma}

\begin{proof}
The proof is similar to that of Lemma~\ref{lemma:weak}, and hence is
omitted.
\end{proof}

We are now in a position to show the completeness of resource proofs
for {\bf PLL}.

\begin{prop}[completeness of resource proofs]\label{prop:maicomplete}
Let $\Gamma \vdash \Delta$ be a sequent in {\bf PLL}.
If $\Gamma \vdash \Delta$ has a proof $\Phi$ in {\bf PLL}, then there
are disjoint sets of Boolean variables $V$ and $W$ such that
$\Gamma . V \vdash \Delta . W$ has a resource proof $R$ and the linear proof
tree corresponding to $R$ is $\Phi$.
\end{prop}

\begin{proof}

By induction on the structure of resource proofs.
Assume that $\Gamma \vdash \Delta$ has a resource proof.

In the base case, the sequent is the conclusion of either the Axiom,
$\perp$L, {\bf 1}R, {\bf 0}L or $\top$R rules, and it is clear that
the result holds in each of these cases.

Hence we assume that the result holds for all provable sequents whose
resource proof is no more than a given size.

Consider the last rule used in the proof. We only give the argument
for the cases $\otimes$L, $\otimes$R, $\with$L, $\with$R, C!L; the
others are similar.

\begin{itemize}

\item[$\otimes$L:]
In this case, we have that
$\Gamma = \phi_1 \otimes \phi_2, \Gamma'$, and there is a proof of
$\Gamma', \phi_1, \phi_2 \vdash \Delta$ which is a subproof of $\Phi$. By the hypothesis, there are disjoint sets of
variables $V$ and $W$ such that there is a resource proof of
$\Gamma'.V, \phi_1[1], \phi_2[1] \vdash \Delta.W$, and
hence there is a resource proof of
$\Gamma'.V, (\phi_1 \otimes \phi_2)[1] \vdash  \Delta.W$, and
clearly the linear proof corresponding to this resource proof is $\Phi$.

\item[$\otimes$R:]
In this case,  we have that
$\Gamma = \Gamma_1, \Gamma_2$, $\Delta = \phi_1 \otimes \phi_2, \Delta_1, \Delta_2$ such that
$\Gamma_1 \vdash \phi_1, \Delta_1$ and
$\Gamma_2 \vdash \phi_2, \Delta_2$
both have proofs in the linear sequent
calculus which are subproofs of $\Phi$. Hence by the hypothesis there are disjoint sets of Boolean
variables $V_1, V_2, W_1, W_2$ such that
$\Gamma_1 . V_1 \vdash \phi_1[1], \Delta_1 . W_1$ and
$\Gamma_2 . V_2 \vdash \phi_2[1], \Delta_2 . W_2$ have resource proofs,
(and moreover the linear proofs corresponding to each resource proof is the appropriate subproof of $\Phi$)
and so by Lemma \ref{lemma:maiweak}, there are closed
resource derivations of
$\Gamma_1 . V_1, \Gamma_2 . \{0\}^n \vdash \phi_1[1], \Delta_1 . W_1, \Delta_2 . \{0\}^n$ and
$\Gamma_1 . \{0\}^n, \Gamma_2 . V_2 \vdash \phi_2[1], \Delta_1 . \{0\}^n, \Delta_2 . W_2$.
Hence there are new disjoint sets of Boolean variables (\ie\ not
occurring anywhere in the above two resource sequents) $V$ and $W$
and a total assignment $I$ of $V \cup W$ such that
$(\Gamma_1 . V_1, \Gamma_2 . V_2). V \vdash \phi_1[1], (\Delta_1 . W_1,  \Delta_2 . W_2). W$ and
$(\Gamma_1 . V_1, \Gamma_2 . V_2). \ov{V} \vdash \phi_2[1], (\Delta_1
. W_1,\Delta_2 . W_2) . \ov{W}$
have resource proofs, and so there is a resource proof of
$\Gamma_1 . V_1, \Gamma_2 . V_2 \vdash (\phi_1 \otimes \phi_2)[1], \Delta_1
. W_1, \Delta_2 . W_2$, \ie\ $\Gamma . V' \vdash (\phi_1 \otimes \phi_2)[1], \Delta . W'$
for some disjoint sets of  Boolean variables $V'$ and $W'$, and clearly the
linear proof corresponding to this resource proof is $\Phi$.

\item[$\with$L:]
In this case, we have that
$\Gamma = \phi_1 \with \phi_2, \Gamma'$, and there is a proof of
$\Gamma', \phi_i \vdash \Delta$ which is a subproof of $\Phi$. Without
loss of generality, let $i = 1$. By the hypothesis, there are disjoint sets of
variables $V$ and $W$ such that there is a resource proof of
$\Gamma'.V, \phi_1[1] \vdash \Delta.W$, and by Lemma~\ref{lemma:maiweak}
there is a resource proof of
$\Gamma'.V, \phi_1[1], \phi_2[0] \vdash \Delta.W$, \ie\ there is a
resource proof of
$\Gamma'.V, \phi_1[x], \phi_2[\ov{x}] \vdash \Delta.W$, giving us a
resource proof of
$\Gamma'.V, (\phi_1 \with \phi_2)[1] \vdash  \Delta.W$, and
clearly the linear proof corresponding to this resource proof is $\Phi$.

\item[$\with$R:]
In this case,  we have that
$\Delta = \phi_1 \with \phi_2, \Delta'$ such that
$\Gamma \vdash \phi_1, \Delta'$ and
$\Gamma \vdash \phi_2, \Delta$
both have proofs in the linear sequent
calculus which are subproofs of $\Phi$. Hence by the hypothesis there are disjoint sets of Boolean
variables $V_1, V_2, W_1, W_2$ such that
$\Gamma . V_1 \vdash \phi_1[1], \Delta . W_1$ and
$\Gamma . V_2 \vdash \phi_2[1], \Delta . W_2$ have resource proofs,
(and moreover the linear proofs corresponding to each resource proof
is the appropriate subproof of $\Phi$).
Now as each of these has the property that the Boolean expression
attached to each formula in the conclusion evaluates to 1, we can
choose $V_1 = V_2$ and $W_1 = W_2$, and hence we have a resource proof
of $\Gamma . V_1 \vdash (\phi_1 \with \phi_2)[1], \Delta . W_1$, and clearly
the linear proof corresponding to this resource proof is $\Phi$.

\item[C!L:]
In this case, we have that
$\Gamma = !\phi, \Gamma'$, and there is a proof of
$\Gamma', !\phi, !\phi \vdash \Delta$ which is a subproof of $\Phi$. By the hypothesis, there are disjoint sets of
variables $V$ and $W$ such that there is a resource proof of
$\Gamma'.V, !\phi[1], !\phi[1] \vdash \Delta.W$, and
hence there is a resource proof of
$\Gamma'.V, !\phi[1] \vdash  \Delta.W$, and
clearly the linear proof corresponding to this resource proof is $\Phi$.
\end{itemize}

\end{proof}

It should be noted that the reason that it is straightforward to
adapt the $\oplus$R and $\with$L rules as above is that the
Boolean constraints to be solved are of the same form as those in \S2,
\ie\ determining the satisfiability of expressions such as
$x.y.\ov{z}$. This may  be thought of as determining the truth of the
existentially quantified expression $\exists{x}.\exists{y}.\exists{z}.(x.y.\ov{z})$.

Now if we were to consider universally quantified expressions as well,
then we arrive at a technique similar to the ``slices'' used in proof
nets for additive rules \cite{girard94}. In particular, we could
re-write the $\otimes$R rule as
$$
\ds\frac{\Gamma \vdash \phi_1[x], \phi_2[\ov{x}], \Delta}{\Gamma \vdash
\phi_1 \otimes \phi_2, \Delta} \otimes {\rm R}
$$
which would result in a universally quantified expression. Hence the
complexity of finding solutions is increased, but for the benefit of
providing an algebraic interpretation of proofs in which the additives
are taken as basic, and the multiplicatives introduced via the Boolean
constraints. This contrasts with proof-nets, in which the additives
are introduced to a multiplicative system by means of the Boolean
expressions. Further development on this point is beyond the scope
of this paper; we will take up this thread in a subsequent paper.

\section{Bunched additives} \label{sec:BI}

We now turn to a different extension of {\bf MLL}, in which the
additives are handle rather differently. Roughly, in the logic
of bunched implications, \BI, there are two implications --- one
multiplicative, and one additive. This necessitates a more complex
structure in the antecedent, as we need to be able to distinguish
additive contexts from multiplicative contexts. However, as we shall
see, the same basic technique as above may be applied to the problem
of distribution of formul{\ae} across multiplicative branches of a
proof.

In \BI, there are multiplicative versions of conjunction and
disjunction (as in linear logic), but also of implication (unlike
linear logic). Hence we have the following two rules for implications
in succedents:

$$
\rightarrow{\rm R}\quad \frac{\Gamma; \phi_1 \vdash \phi_2}
{\Gamma \vdash \phi_1 \rightarrow \phi_2}
\qquad
\bimp{\rm R}\quad  \frac{\Gamma, \phi_1 \vdash \phi_2}
{\Gamma \vdash \phi_1 \bimp \phi_2}
$$

These rules thus show that implication is intimately associated
with the way in which antecedents are formed. In particular, because
of the existence of the two implication rules above, it is necessary
to have two constructors for antecedents: one multiplicative, written
as ``,'' and logically equivalent to \BI's multiplicative conjunction,
$*$,  and one additive, written as ``;'' and logically equivalent to
\BI's additive conjunction, $\wedge$.\footnote{The underlying semantic
structure of \BI's proofs, which can be represented as the terms of
a simply-typed $\lambda$-calculus, is given by a \emph{bi-cartesian doubly
closed category} \cite{OP99,pym99lics,Pym00Mono}, which carries two closed
structures, one symmetric monoidal and one bi-cartesian. There are
many examples of such categories \cite{OP99,pym99lics,Pym00Mono}.}

Thus antedecents in \BI\ are not simply finite sequences, as in linear
logic, but finite trees, with the leaf nodes being formul{\ae}, and the
internal nodes denoted by either ``,'' or ``;''~, and are referred to as
\emph{bunches}.\footnote{A multiple-conclusioned version of \BI, with
disjunctive bunched structure in succedents, may also be formulated
\cite{Pym00Mono}. For simplcity, we restrict our attention here to the
single-conclusioned case.} The grammar of bunches is given in
Figure~\ref{fig:bunches}.

\begin{figure}
\hrule\hspace{5mm}

$$
\begin{array}{lcrlr}
\Gamma & ::=& & \phi & \mbox{propositional assumption}\\
&& \mid&
\emptyset_m & \mbox{multiplicative unit}\\
&&\mid &
\Gamma ,  \Gamma
& \mbox{multiplicative combination}\\
&& \mid&
\emptyset_a & \mbox{additive unit}\\
&&\mid &
\Gamma ; \Gamma
& \mbox{additive combination}\\
\end{array}
$$

\hspace{5mm}
\hrule
\caption{Bunches}
\label{fig:bunches}
\vspace{0.1in}
\end{figure}

\begin{figure}
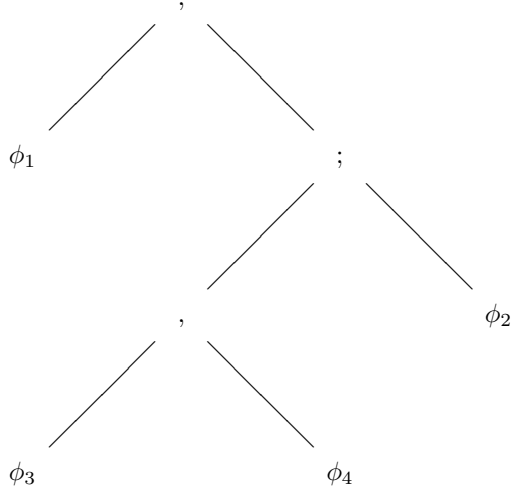

\hrule\hspace{5mm}

\begin{diagram}
   &            &         & , &         &           &         &          & & & \\
   &            & \ruLine &   & \luLine &           &         &          & & & \\
   & \phi_{1}   &         &   &         & ;         &         &          & & & \\
   &            &         &   & \ruLine &           & \luLine &          & & & \\
   &            &         & , &         &           &         & \phi_{2} & & & \\
   &            & \ruLine &   & \luLine &           &         &          & & & \\
   & \phi_{3}   &         &   &         & \phi_{4}  &         &          & & & \\
\end{diagram}

\hspace{5mm}
\hrule
\caption{An example bunch}
\label{fig:example-bunch}
\vspace{0.1in}
\end{figure}

For example, the bunch $\phi_{1} , ( (\phi_{3} , \phi_{4}) ; \phi_{2})$ may be drawn as in
Figure~\ref{fig:example-bunch}.

The main point of the definition of bunches is that ``;'' admits the
structural properties of Weakening and Contraction, whereas ``,'' does
not: this distinction allows the correct formulation of the two
implications. Bunches may be represented using lists of lists, \etc,
as described in \cite{Read88}. We write $\Gamma(\Delta)$, and refer
to $\Delta$ as a \emph{sub-bunch} of $\Gamma$, for a bunch $\Gamma$ in
which $\Delta$ appears as a sub-tree and write $\Gamma[\Delta'/\Delta]$
for $\Gamma$ with $\Delta$ replaced by $\Delta'$. We write $\Gamma(-)$
to denote a bunch $\Gamma$ which is incomplete and which may be
completed by placing a bunch in its hole, and will use this notation
to refer to that part of $\Gamma(\Delta)$ which is not part of $\Delta$.
We require that ``,'' and ``;'' be commutative monoids, giving rise to
the coherent equivalence,\index{coherent equivalence} $\Gamma \equiv \Gamma'$,
as follows:\footnote{The multiplicative operation, ``,'', need not, in general,
be commutative and, indeed, such non-commutative versions of \BI\ are
possible \cite{Pym00Mono}.}

\medskip

\noindent {\sc Coherent Equivalence:}\/ $\Gamma \equiv \Gamma'$
\begin{enumerate}
\item [1] Commutative monoid equations for
$\emptyset_a$ and ``;''.
\item [2] Commutative monoid equations for
$\emptyset_m$ and ``,''.
\item [3] Congruence: if $\Delta \equiv \Delta'$ then
$\Gamma(\Delta) \equiv \Gamma(\Delta')$.
\end{enumerate} \index{coherent equivalence}
Note that ``;'' and ``,'' do not distribute over one another. We use
$=$ for syntactic identity of bunches.


\BI\ may also be understood semantically \cite{OP99,pym99lics,Pym00Mono}.
The key structure is that of a \emph{preordered monoid},
${\cal M} = (M,\cdot,e,\sqsubseteq)$ which
provides the worlds for a Kripke-style forcing semantics. The additives
receive a semantics in the usual intuitionistic way, whereas the
multiplicatives receive a semantics in the style of Kripke, Urquhart
and Pym \cite{Kripke65,Urquhart72,OP99,pym99lics,Pym00Mono}:
$$
\begin{array}{rcl}
m \models \phi \mimp \psi & \mbox{\rm iff} &
  \mbox{\rm for all $n$ such that $n \models \phi$, $m \cdot n \models \psi$}\\
m \models \phi *     \psi & \mbox{\rm iff} &
  \mbox{\rm there exist $n$ and $n'$, where $m \sqsubseteq n \cdot n'$, such that
$n \models \phi$ and $n' \models \psi$.}\\
\end{array}
$$
Categorically, such a semantics can be formulated in doubly closed categories
of the form ${\bf Set}^{C}$, where $C$ is a small symmetric monoidal
\cite{OP99,pym99lics,Pym00Mono}. This semantics may be interpreted
as being a model of the  notion of \emph{resource}: it may be argued that
the following are a basic set of assumptions about resources:
\begin{itemize}
\item There should be a \emph{zero} resource; call it $e$;
\item The should be a \emph{combination} of resources;  call it $\cdot$;
\item The should be a a \emph{comparison} of resources; call it $\sqsubseteq$.
\end{itemize}
The assumptions lead to the forcing semantics based on preordered monoids. The
leading characteristic of this resource semantics is the
\emph{sharing interpretation}:
we can analyse the sharing and privacy of the resources accessed by co-existing
computations. The topics, including soundness and completeness theorems and
computationally natural concrete models, are
discussed fully in \cite{OP99,pym99lics,Pym00Mono,OPY}. The logical view
of resources sketched here should be contrasted with that which is available
for linear logic, in which a proposition is interpreted directly as a
resource and its number of uses is counted.

For further information on \BI, the reader is referred to
\cite{OP99,pym99lics,Pym00Mono}. A good summary of the use
of bunched structure in substructural logic may be found in
\cite{Restall2000,Read88}.

Below is (a version of) the sequent calculus {\bf LBI} for propositional
{\bf BI} \cite{OP99,pym99lics,Pym00Mono}.

\begin{define} The calculus {\bf LBI} may be defined as follows:

\begin{tabular}{llll}
& & & \\

Axiom & $\ds \frac{}{\phi \vdash \phi}$ &
{\rm E} & $\ds \frac{\Gamma \vdash \phi}{\Delta \vdash \phi}$ where $\Gamma \equiv \Delta$ \\ & & & \\

{\rm W} & $\ds \frac{\Gamma(\Delta) \vdash \phi}{\Gamma(\Delta; \Delta') \vdash \phi}$  &
{\rm C} & $\ds \frac{\Gamma(\Delta; \Delta) \vdash \phi}{\Gamma(\Delta) \vdash \phi}$ \\
& & & \\

$I${\rm L} & $\ds \frac{\Gamma(\emptyset_m) \vdash \phi}{\Gamma(I) \vdash \phi}$ &
$I${\rm R} & $\ds \frac{}{\emptyset_m \vdash I}$ \\
& & & \\

$1${\rm L} & $\ds \frac{\Gamma(\emptyset_a) \vdash \phi}{\Gamma(1) \vdash \phi}$  &
$1${\rm R} & $\ds \frac{}{\emptyset_a \vdash 1}$ \\
& & & \\

$\perp${\rm L} & $\ds \frac{}{\perp\; \vdash \phi}$ \\& & &  \\

$\bimp${\rm L} & $\ds \frac{\Gamma \vdash \phi_1 \hspace{0.2in} \Delta(\Delta',\phi_2) \vdash \phi}
                {\Delta(\Delta', \Gamma, \phi_1 \bimp \phi_2) \vdash \phi}$ &
$\bimp${\rm R} & $\ds \frac{\Gamma, \phi_1 \vdash \phi_2}{\Gamma \vdash \phi_1 \bimp
\phi_2}$ \\
& & & \\

$*${\rm L} & $\ds \frac{\Gamma(\phi_1, \phi_2)  \vdash \phi}{\Gamma(\phi_1 * \phi_2) \vdash \phi}$ &
$*${\rm R} & $\ds \frac{\Gamma \vdash \phi_1 \hspace{0.2in} \Delta \vdash \phi_2}
                {\Gamma, \Delta \vdash \phi_1 * \phi_2}$ \\
& & & \\

$\rightarrow${\rm L} & $\ds \frac{\Gamma \vdash \phi_1 \hspace{0.2in} \Delta(\Gamma; \phi_2) \vdash \phi}
                {\Delta(\Gamma; \phi_1 \rightarrow \phi_2) \vdash \phi}$ &
$\rightarrow${\rm R} & $\ds \frac{\Gamma; \phi_1 \vdash \phi_2}{\Gamma \vdash \phi_1
\rightarrow \phi_2}$ \\
& & & \\

$\land${\rm L} & $\ds \frac{\Gamma(\phi_1; \phi_2)  \vdash \phi}{\Gamma(\phi_1 \land \phi_2) \vdash \phi}$ &
$\land${\rm R} & $\ds \frac{\Gamma \vdash \phi_1 \hspace{0.2in} \Gamma \vdash \phi_2}
                {\Gamma \vdash \phi_1 \land \phi_2}$ \\
& & & \\

$\lor${\rm L} & $\ds \frac{\Gamma(\phi_1) \vdash \phi \hspace{0.2in} \Gamma(\phi_2) \vdash \phi}
                  {\Gamma(\phi_1 \lor \phi_2) \vdash \phi}$ &
$\lor${\rm R} & $\ds \frac{\Gamma \vdash \phi_i}{\Gamma \vdash \phi_1 \lor \phi_2}$ \\
& & & \\

\end{tabular}

\noindent We will refer to the Axiom, $I$R, $1$R and $\perp$L rules as \emph{leaf} rules.

\end{define}

Note that, since {\bf LBI} admits Cut-elimination \cite{Pym00Mono}, we omit the Cut rule.
Also, we have used the additive version of the rules $\land$R, $\rightarrow$L and
$\lor$L rules (see \cite{OP99,pym99lics,Pym00Mono} for details). This is done
because the derivability of these rules in {\bf LBI} may  be interpreted as
showing that these rules need not introduce any new distribution problems.
Note also that, as in \cite{OP99,pym99lics,Pym00Mono}, we could have used a
similarly additive form of the Axiom rule, \emph{viz.}:

\begin{center}
Axiom $\ds \frac{}{\phi ; \Gamma \vdash \phi}$
\end{center}

However, we have chosen not to do this as it unnecessarily complicates
the structure of the antecedent.  Recall that the basic technique is
to relax the multiplicative rules so that ``extra'' formul{\ae} are present
in the sequent, and then the constraints are used to determine the
distribution of all formul{\ae} across multiplicative branches. In the
linear case, as antecedents are multiplicative, there is no issue.  In
the {\bf BI} case, the antecedent could be additive, as in the above
rule. The resource version of the Axiom rule would then be

\begin{center}
Axiom $\ds \frac{\forall e \in {\rm exp}(\Delta) \; e = 0}
             {(\phi[e_1] ; \Gamma), \Delta \vdash \phi}$ $e_1 = 1$,
\end{center}

\noindent which seems unnecessarily complex, especially as the rules of
weakening and contraction are still required. Hence we use the original
form of the Axiom rule but the additive versions of the $\land$R,
$\rightarrow$L and $\lor$L rules.

Now the only rules which require attention are the leaf rules
leaves of the tree (Axiom, $I$R, $\perp$L, $1$R) and $\bimp$L (as
$*$R is precisely the same rule as for linear logic). We reformulate
the leaf rules as follows:

\begin{center}
\begin{tabular}{ll}
& \\
Axiom $\ds \frac{\forall e \in {\rm exp}(\Delta) \; e = 0}
                {\phi[e_1], \Delta \vdash \phi}$ $e_1 = 1$ &

$\perp$ðL $\ds \frac{\forall e \in {\rm exp}(\Delta) \; e = 0}
                   {\perp[e_1], \Delta \vdash \phi}$ $e_1 = 1$ \\ & \\

$I$ðR $\ds \frac{\forall e \in {\rm exp}(\Delta) \; e = 0}
               {\emptyset_m[e_1], \Delta \vdash I}$ $e_1 = 1$ &

$1$ðR $\ds \frac{\forall e \in {\rm exp}(\Delta) \; e = 0}
               {\emptyset_a[e_1], \Delta \vdash 1}$ $e_1 = 1$ \\& \\
\end{tabular}
\end{center}

\noindent Note that as {\bf LBI} is single-conclusioned, there is no need to
consider distribution on the right of $\vdash$, and hence the formul{\ae}
in the succedents never require a Boolean expression.  Otherwise,
these are little different from similar rules in linear logic.

The only remaining rule is $\bimp$L. Here the main point to note is
that we need to take account of the bunch structure, but otherwise
this is handled similarly to the $\limp$L rule in linear logic.
The new form of the rule is then
$$
\frac{\Gamma.V \vdash \phi_1 \hspace{0.2in} \Delta(\Gamma.\ov{V}, \phi_2[e]) \vdash \phi}
          {\Delta(\Gamma, (\phi_1 \bimp \phi_2)[e]) \vdash \phi} \; e = 1
$$
where $\Gamma, \phi_1 \bimp \phi_2$ is the maximal multiplicative super-bunch
of $\phi_1 \bimp \phi_2$ in $\Delta(\Gamma, \phi_1 \bimp \phi_2)$. For example, given
the sequent

\begin{center}
$p, (q; (r,s,t \bimp u)) \vdash v$
\end{center}

\noindent  we get the following inference step:
$$
\frac{r[x_1], s[x_2] \vdash t \hspace{0.2in} p, (q; (r[\ov{x_1}], s[\ov{x_2}], u[1])) \vdash v}
          {p, (q; (r,s,(t \bimp u)[1])) \vdash v}
$$
Here $\Gamma = \{r,s\}$ and not $\{r\}$ or $\{s\}$. We formalize this notion below.

\begin{define}
Let $\Delta(\Gamma)$ be a bunch.
Then the \emph{maximal multiplicative super-bunch} of $\Gamma$
in $\Delta$ is defined as follows:

\begin{itemize}
\item If $\Delta(\Gamma) = \Gamma$ or the parent of $\Gamma$ in
$\Delta$ is ``{\rm ;}'',
then the maximal multiplicative super-bunch of $\Gamma$ in $\Delta$
is just $\Gamma$.

\item If the parent of $\Gamma$ in $\Delta$ is ``{\rm ,}'',
then the maximal multiplicative super-bunch of $\Gamma$ in $\Delta$
is the maximal multiplicative super-bunch of the parent of $\Gamma$ in
$\Delta$.
\end{itemize}
\end{define}

Note that in this case we have to attach expressions to bunches. For example,
consider the provable sequent $r, (s;t) \vdash r * s$.
A resource derivation of this sequent would be

$$
\infer[*{\rm R}]{r, (s;t) \vdash r * s}
  {
    r[x], (s;t)[y] \vdash r
    &
    \infer[{\rm W}]{r[\ov{x}], (s;t)[\ov{y}] \vdash s}{r[\ov{x}], s[\ov{y}] \vdash s}
  }
$$

from which we get $x = 1$, $y = 0$, corresponding to the sequent proof below.

$$
\infer[*{\rm R}]{r, (s;t) \vdash r * s}
  {
    r \vdash r
    &
    \infer[{\rm W}]{s;t \vdash s}{s \vdash s}
  }
$$

Hence we need to attach expressions to bunches. However, it will not
necessarily be correct to simply attach the expression to every
formula in the bunch, as a multiplicative bunch may  contain an
additive sub-bunch. For example, in the sequent above, we wish to
distribute the two sub-bunches $r$ and $(s;t)$ across the two
multiplicative branches, and not the three sub-bunches $r$, $s$ and
$t$.

Hence we need to define the operation $\Gamma . V$
where $\Gamma$ is a bunch so that it respects this multiplicative
structure.

\begin{define}
Let $\Gamma$ be a bunch and $V$ a set of distinct variables. We define
the operation $\Gamma . V$ as follows:

\begin{itemize}
\item If $\Gamma = \phi$, then $|V| = 1$ and $\Gamma . V = \phi[x]$
\item If $\Gamma = (\Gamma';\Delta)$, then
$|V| = 1$ and $\Gamma . V = (\Gamma';\Delta)[x]$
\item If $\Gamma = (\Gamma',\Delta)$, then
$\Gamma . V = (\Gamma'.V_1, \Delta.V_2)$
where $V_1$, $V_2$ are disjoint sets of variables such that $V = V_1 \cup V_2$.
\end{itemize}

\end{define}

Thus we have that $r, (s;t) . \{x,y\} = r[x], (s;t)[y]$.

Note that the linear version of this operation may  be thought of as a
bunch in which the only constructor is {\bf ``,''}.

Note also that in the additive implication it is also useful to
identify the maximal \emph{additive} super-bunch of the principal
formula. Doing so allows us to show that there is indeed no
distribution problem arising from the additive form of the
$\rightarrow$L rule, as then $\Gamma$ and $\Delta$ may easily be
determined from the rule application.

For example, given the sequent

\begin{center}
$q, (p_3; p_1; p_1 \rightarrow p_2) \vdash q * p_2$
\end{center}

it would seem appropriate to apply the $\rightarrow$L rule as follows

$$
\infer[\rightarrow {\rm L}]{q, (p_3; p_1; p_1 \rightarrow p_2) \vdash
q * p_2}
{
    \infer[W]{p_3; p_1 \vdash p_1}{p_1 \vdash p_1}
    &
    \infer[*R]{q, (p_3; p_1; p_2) \vdash q * p_2}
    {
        q \vdash q
        &
        \infer[W]{p_3; p_1; p_2 \vdash p_2}{p_2 \vdash p_2}
    }
}
$$

rather than

$$
\infer[\rightarrow {\rm L}]{q, (p_3; p_1; p_1 \rightarrow p_2) \vdash
q * p_2}
{
    p_1 \vdash p_1
    &
    \infer[*R]{q, (p_3; p_1; p_2) \vdash q * p_2}
    {
        q \vdash q
        &
        \infer[W]{p_3; p_1; p_2 \vdash p_2}{p_2 \vdash p_2}
    }
}
$$

The former application avoids having to determine where $p_3$ will be
used by sending it to each branch and deleting it (via the W rule) as
required. However, we do not pursue this direction in this paper, as
it has no effect on the issues of distribution.


We get the following rules:

\begin{define} \label{def:rulesBI}
We define the following \emph{sequent calculus with constraints} for
\BI:

\begin{tabular}{llll}
& \\
{\rm Axiom} & $\ds \frac{\forall e \in {\rm exp}(\Delta) \; e = 0}
                {\phi[e_1], \Delta \vdash \phi}$ $e_1 = 1$ &
{\rm E} & $\ds \frac{\Gamma[e] \vdash \phi}{\Delta[e] \vdash \phi}$ $e = 1$ where $\Gamma \equiv \Delta$ \\
& & & \\

{\rm W} & $\ds \frac{\Gamma(\Delta[e]) \vdash \phi}{\Gamma((\Delta; \Delta')[e]) \vdash \phi}$ $e = 1$ &
{\rm C} & $\ds \frac{\Gamma(\Delta[e]; \Delta[e]) \vdash
\phi}{\Gamma(\Delta[e]) \vdash \phi}$ $e = 1$ \\
& & & \\

$I${\rm L} & $\ds \frac{\Gamma(\emptyset_m[e]) \vdash \phi}{\Gamma(I[e]) \vdash \phi}$ $e = 1$ &
$I${\rm R} & $\ds \frac{\forall e \in {\rm exp}(\Delta) \; e = 0}
               {\emptyset_m[e_1], \Delta \vdash I}$ $e_1 = 1$\\
& & & \\

$1${\rm L} & $\ds \frac{\Gamma(\emptyset_a[e]) \vdash \phi}{\Gamma(1[e]) \vdash \phi}$ $e = 1$ &
$1${\rm R} & $\ds \frac{\forall e \in {\rm exp}(\Delta) \; e = 0}
               {\emptyset_a[e_1], \Delta \vdash 1}$ $e_1 = 1$\\
& & & \\

$\perp${\rm L} & $\ds \frac{\forall e \in {\rm exp}(\Delta) \; e = 0}
                   {\perp[e_1], \Delta \vdash \phi}$ $e_1 = 1$\\
& & & \\

$\bimp${\rm L} & $\ds \frac{\Gamma.V \vdash \phi_1 \hspace{0.2in} \Delta(\Gamma.\ov{V}, \phi_2[e]) \vdash \phi)}
                  {\Delta(\Gamma, \phi_1 \bimp \phi_2[e]) \vdash \phi}$ $e = 1$&
$\bimp${\rm R} & $\ds \frac{\Gamma, \phi_1[e] \vdash \phi_2}{\Gamma \vdash \phi_1
\bimp \phi_2}$ $e = 1$ \\
& & & \\

$*${\rm L} & $\ds \frac{\Gamma(\phi_1[e], \phi_2[e])  \vdash \phi}{\Gamma(\phi_1 * \phi_2[e]) \vdash \phi}$ $e = 1$ &
$*${\rm R} & $\ds \frac{\Gamma.V \vdash \phi_1 \hspace{0.2in} \Gamma.\ov{V} \vdash \phi_2}
                {\Gamma \vdash \phi_1 * \phi_2}$ \\
& & & \\

$\rightarrow${\rm L} & $\ds \frac{\Gamma \vdash \phi_1 \hspace{0.2in} \Delta(\Gamma; \phi_2[e]) \vdash \phi}
                {\Delta(\Gamma; \phi_1 \rightarrow \phi_2[e]) \vdash \phi}$ $e = 1$ &
$\rightarrow${\rm R} & $\ds \frac{\Gamma; \phi_1[e] \vdash \phi_2}{\Gamma \vdash
\phi_1 \rightarrow \phi_2}$ $e = 1$ \\
& & & \\

$\land${\rm L} & $\ds \frac{\Gamma(\phi_1[e]; \phi_2[e])  \vdash \phi}{\Gamma(\phi_1 \land \phi_2[e]) \vdash \phi}$ $e = 1$ &
$\land${\rm R} & $\ds \frac{\Gamma \vdash \phi_1 \hspace{0.2in} \Gamma \vdash \phi_2}
                {\Gamma \vdash \phi_1 \land \phi_2}$  \\
& & & \\

$\lor${\rm L} & $\ds \frac{\Gamma(\phi_1[e]) \vdash \phi \hspace{0.2in} \Gamma(\phi_2[e]) \vdash \phi}{\Gamma(\phi_1 \lor \phi_2[e]) \vdash \phi}$ $e = 1$ &
$\lor${\rm R} & $\ds \frac{\Gamma \vdash \phi_i}{\Gamma \vdash \phi_1 \lor \phi_2}$
\\
& & & \\
\end{tabular}

\noindent   where the $\bimp${\rm L} rule has the side condition that
$\Gamma, \phi_1 \bimp \phi_2$ is the maximal multiplicative super-bunch
containing $\phi_1 \bimp \phi_2$,
and the rules $\bimp${\rm L} and $*${\rm R} have the side condition that $V$ is a
set of Boolean variables not occurring in $\Gamma$, $\Delta$, $\phi_1$,
$\phi_2$ or $\phi$.
\end{define}

We are now in a position to define resource derivations for
propositional \BI. This is actually the same definition as
Definition~\ref{def:resource-derivation}.

\begin{define}
A \emph{resource derivation} is a tree regulated by the
rules of the resource calculus in which each formula of the endsequent
is assigned a distinct Boolean variable, together with a (possibly partial) assignment
of the Boolean variables appearing in the derivation.

A resource derivation is \emph{total} if its assignment of the
Boolean variables is total. Otherwise, the resource derivation is {\em
partial}. A resource derivation is \emph{closed} if all of the leaves of the
proof tree are leaf rules.
A \emph{resource proof} is a total, closed resource derivation in
which all the Boolean variables in the endsequent
are assigned the value $1$.
\end{define}

As above, it is then straightforward to recover an \LBI\ proof from a
resource proof.

\begin{define}
Let $R$ be a total resource derivation, with proof tree $T$ and
Boolean assignment $I$. The {\bf LBI} \emph{proof tree corresponding
to R} is the proof tree obtained by deleting from $T$ all formul{\ae}
whose Boolean expression evaluates to $0$ under $I$.
\end{define}

For example, consider the resource derivation below.

{\small
$$
   \infer[\mimp {\rm L}]{(r, (p;t), p \bimp q[1]); s \vdash q * r}
    {
      \infer[W]{r[x_1], (p;t)[x_2] \vdash p}{r[x_1], p[x_2] \vdash p}
      &
      \infer[W]{(r[\ov{x_1}], (p;t)[\ov{x_2}], q);s \vdash q * r }
      {
        \infer[* {\rm R}]{r[\ov{x_1}], (p;t)[\ov{x_2}], q \vdash q * r }
        {
           r[\ov{x_1}.y_1], (p;t)[\ov{x_2}.y_2], q[y_3] \vdash q
           &
           r[\ov{x_1}.\ov{y_1}], (p;t)[\ov{x_2}.\ov{y_2}], q[\ov{y_3}] \vdash r
        }
      }
    }
$$
}

Solving the equations gives us

\begin{center}
$x_1 = 0, x_2 = 1, y_1 = 0, y_3 = 1$
\end{center}

which in turn gives us the resource proof:

{\small
$$
   \infer[\mimp {\rm L}]{(r, (p;t), p \bimp q[1]); s \vdash q * r}
    {
      \infer[W]{r[0], (p;t)[1] \vdash p}{r[0], p[1] \vdash p}
      &
      \infer[W]{(r[1], (p;t)[0], q);s \vdash q * r }
      {
         \infer[* {\rm R}]{r[1], (p;t)[0], q \vdash q * r }
         {
           r[0], (p;t)[0], q[1] \vdash q
           &
           r[1], (p;t)[0], q[0] \vdash r
         }
      }
    }
$$
}

The corresponding {\bf LBI} proof is then:

{\small
$$
   \infer[\mimp {\rm L}]{(r, (p;t), p \bimp q); s \vdash q * r}
    {
      \infer[W]{(p;t) \vdash p}{p \vdash p}
      &
      \infer[W]{(r, q);s \vdash q * r }
      {
         \infer[* {\rm R}]{r, q \vdash q * r }
         {
           q \vdash q
           &
           r \vdash r
         }
      }
    }
$$
}

Note that this proof could be part of a larger proof, such as the one
below, but no new constraints are introduced by the larger context.

{\small
$$
\infer[\land {\rm R}]{(r, (p;t), p \bimp q); s \vdash (q * r) \land s}
{
   \infer[\mimp {\rm L}]{(r, (p;t), p \bimp q); s \vdash q * r}
    {
      \infer[W]{(p;t) \vdash p}{p \vdash p}
      &
      \infer[W]{(r, q);s \vdash q * r }
      {
         \infer[* {\rm R}]{r, q \vdash q * r }
         {
           q \vdash q
           &
           r \vdash r
         }
      }
    }
   &
   \infer[W]{(r, (p;t), p \bimp q); s \vdash s}{s \vdash s}
}
$$
}

It should be noted that the bunch structure may also simplify some
matters. For example, when two variables share the same expression in
a multiplicative context, such as the bunch $r[x], s[x]$, it is not
hard to see that this is equivalent to $(r,s)[x]$, which is a simpler
representation of the same information. It may happen that this latter
structure requires a decomposition because of  a need to split this
multiplicative bunch, in which case we would revert to the former form
(and presumably add extra variables to each expression,
\eg\ $r[x.y], s[x.z]$). Thus the bunch structure introduces some
further dynamics in the ``tagging'' of formul{\ae}.

It should also be noted that this tagging is transparent to coherent
equivalence. In particular, it should be clear that if $\Gamma \equiv
\Delta$, then $\Gamma(\Delta'[0]) \equiv \Delta(\Delta'[0])$ and
$\Gamma(\Delta'[1]) \equiv \Delta(\Delta'[1])$.
It should also be clear that (by abuse of notation)
$\Gamma.V(\Delta[0]) = \Gamma(\Delta[0]).V$.

We are now in a position to show the soundness of resource proofs for
\BI.

\begin{prop}[soundness of resource proofs]\label{prop:bisound}
Let $\Gamma \vdash \phi$ be a resource sequent in \BI.
If $\Gamma \vdash \phi$ has a resource proof $R$ with Boolean
assignment $I$, then the {\bf LBI}
proof tree corresponding to $R$ is an {\bf LBI} proof of
$\Gamma[I]^1 \vdash \phi$.
\end{prop}

\begin{proof}

We proceed by induction on the size of the resource derivation.

In the base case, the resource proof consists of just one of the leaf
rules. Hence there are four cases, of which we only give the argument
for Axiom, the others being similar.


 In this case, the endsequent of the resource proof is just
 $\phi[e_1], \Delta \vdash \phi$
 with $e_1 = 1$ and $\forall e \in {\rm exp}(\Delta) e = 0$.
 Hence $\Gamma[I]^1 = \phi$ and
 the {\bf LBI} proof corresponding to this resource proof is just
 $\phi \vdash \phi$.

%
%
%

Hence we assume that the result holds for all proofs of no more than a given size.

There are numerous cases, of which we only give the argument for
$\bimp$L, $\bimp$R, $*$L, $*$R and $\rightarrow$L, the others being similar.

\begin{itemize}

\item[$\bimp$L:]
In this case, the endsequent is $\Delta(\Gamma, \phi_1 \bimp \phi_2[e]) \vdash \phi$
where $e = 1$, and the premisses are
$\Gamma.V \vdash \phi_1$ and $\Delta(\Gamma.\ov{V}, \phi_2[e]) \vdash \phi)$.
Now as the endsequent has a resource proof,
there is a Boolean assignment $I$ of $V$ such that both premisses have resource proofs.
Hence by the induction hypothesis we have that
the {\bf LBI} proof tree corresponding to the resource derivation of
$\Gamma.V \vdash \phi_1$ is an {\bf LBI} proof of $\Gamma.V[I]^1 \vdash \phi_1$, and
the {\bf LBI} proof tree corresponding to the resource derivation of
$\Delta(\Gamma.\ov{V}, \phi_2[1]) \vdash \phi$  is an {\bf LBI} proof of
$\Delta(\Gamma.\ov{V}, \phi_2[1])[I]^1 \vdash \phi_1$.
Now as $\Gamma, \phi_1 \bimp \phi_2$ is the maximal multiplicative super-bunch of
$\phi_1 \bimp \phi_2$ in $\Delta$, and as $I$ is a total assignment of Boolean
variables,
we have that $\Gamma.V[I]^1$ = $\Gamma_1$ and $\Gamma.\ov{V}[I]^1$ =
$\Gamma_2$ where $\Gamma_1, \Gamma_2 = \Gamma$,
and so the {\bf LBI} proof tree corresponding
to the resource derivation of $\Delta(\Gamma, \phi_1 \bimp \phi_2[1]) \vdash \phi$
is an {\bf LBI} proof of $\Delta(\Gamma, \phi_1 \bimp \phi_2[1])[I]^1 \vdash \phi$.

\item[$\bimp$R:]
In this case, the endsequent is $\Gamma \vdash \phi_1 \bimp \phi_2$, and the premiss is
$\Gamma, \phi_1[1] \vdash \phi_2$.
Hence by the induction hypothesis we have that
the {\bf LBI} proof tree corresponding to the resource derivation of
$\Gamma, \phi_1[1] \vdash \phi_2$
is an {\bf LBI} proof of $\Gamma, \phi_1[1][I]^1 \vdash \phi_2$,
and hence the {\bf LBI} proof tree corresponding to the resource derivation of
$\Gamma \vdash \phi_1 \bimp \phi_2$ is an {\bf LBI} proof of $\Gamma[I]^1 \vdash \phi_1 \bimp \phi_2$.

\item[$*$L:]
In this case, the endsequent is $\Gamma(\phi_1 * \phi_2[e]) \vdash \phi$ where $e = 1$,
and the premiss is $\Gamma(\phi_1[e], \phi_2[e]) \vdash \phi$.
Hence by the induction hypothesis we have that
the {\bf LBI} proof tree corresponding to the resource derivation of
$\Gamma(\phi_1[1], \phi_2[1]) \vdash \phi$
is an {\bf LBI} proof of $\Gamma(\phi_1, \phi_2)[I]^1 \vdash \phi$,
and hence the {\bf LBI} proof tree corresponding to the resource derivation of
$\Gamma(\phi_1 * \phi_2[1]) \vdash \phi$ is an {\bf LBI} proof of $\Gamma(\phi_1 * \phi_2)[I]^1 \vdash \phi$.

\item[$*$R:]
In this case, the endsequent is $\Gamma \vdash \phi_1 * \phi_2$, and
the premisses are $\Gamma.V \vdash \phi_1$ and $\Gamma.\ov{V} \vdash
\phi_2$, and so there is a Boolean assignment $I$ such that both
premisses have a resource-proof.  Hence by the induction hypothesis we
have that the {\bf LBI} proof tree corresponding to the resource
derivation of $\Gamma.V \vdash \phi_1$ is an {\bf LBI} proof of
$\Gamma.V[I]^1 \vdash \phi_1$, and that the {\bf LBI} proof tree
corresponding to the resource derivation of $\Gamma.\ov{V} \vdash
\phi_2$ is an {\bf LBI} proof of $\Gamma.\ov{V}[I]^1 \vdash
\phi_2$.
As $I$ is a total assignment of Boolean variables, we have that
$\Gamma.V[I]^1$ = $\Gamma_1$ and $\Gamma.\ov{V}[I]^1$ = $\Gamma_2$
where $\Gamma_1, \Gamma_2 = \Gamma$, and so
the {\bf LBI} proof tree corresponding to the resource derivation of
$\Gamma \vdash \phi_1 * \phi_2$
is an {\bf LBI} proof of $\Gamma \vdash \phi_1 * \phi_2$.

\item[$\rightarrow$L:]
In this case, the endsequent is $\Gamma(\phi_1 \rightarrow \phi_2[e])
\vdash \phi$ where $e = 1$, and the premisses are $\Gamma \vdash
\phi_1$ and $\Gamma(\phi_2[e]) \vdash \phi$.  Hence by the induction
hypothesis we have that the {\bf LBI} proof tree corresponding to the
resource derivation of $\Gamma \vdash \phi_1$ is an {\bf LBI} proof of
$\Gamma[I]^1 \vdash \phi_1$, and the {\bf LBI} proof tree
corresponding to the resource derivation of $\Gamma(\phi_2) \vdash
\phi$ is an {\bf LBI} proof of $\Gamma(\phi_2)[I]^1 \vdash \phi$, and
hence the {\bf LBI} proof tree corresponding to the resource
derivation of $\Gamma(\phi_1 \rightarrow \phi_2[1]) \vdash \phi$ is an
{\bf LBI} proof of $\Gamma(\phi_1 \rightarrow \phi_2)[I]^1 \vdash
\phi$.

\end{itemize}
\end{proof}

As for {\bf MLL} and {\bf PLL}, we require the following lemma:

\begin{lemma} \label{lemma:biweak}
Let $\Gamma \vdash \phi$ be a resource sequent in \BI.
If $\Gamma \vdash \phi$ has a closed resource derivation, then
for any bunch $\Delta'$ we have that
$\Gamma(\Delta'[0]) \vdash \phi$ also has a closed resource derivation.
\end{lemma}

\begin{proof}

We proceed by induction on the size of the resource derivation.

In the base case, the resource derivation consists of just one of the leaf
rules. Hence there are four cases, of which we only give the argument
for Axiom, the others being similar.

%

In this case, $\Gamma = \phi, \Delta$ and the endsequent of the resource derivation is just
$\phi[e_1], \Delta \vdash \phi$ with $e_1 = 1$ and $\forall e \in {\rm exp}(\Delta) e = 0$.
Hence it is clear that there is also a closed resource derivation of
$\Gamma(\Delta'[0]) \vdash \phi$.

%
%
%

Hence we assume that the result holds for all proofs of no more than a given size.

There are numerous cases, of which we only give the argument for
$\bimp$L, $\bimp$R, $*$L, $*$R and $\rightarrow$L, the others
being similar.

\begin{itemize}

\item[$\bimp$L:] In this case, the endsequent is
$\Delta(\Gamma, \phi_1 \bimp \phi_2[e]) \vdash \phi$ where $e = 1$, and the premisses are
$\Gamma.V \vdash \phi_1$ and $\Delta(\Gamma.\ov{V}, \phi_2[e]) \vdash \phi)$.
Hence by the hypothesis we have that $\Gamma.V(\Delta'[0]) \vdash \phi_1$
and $\Delta(\Gamma(\Delta'[0]).\ov{V}, \phi_2[e])) \vdash \phi)$ both have
closed resource derivations (recall that the position of the bunch $\Delta'$ may be arbitrary),
and so
$\Delta(\Gamma(\Delta'[0]), \phi_1 \bimp \phi_2[e]) \vdash \phi$ has a closed resource derivation.

\item[$\bimp$R:]
In this case, the endsequent is $\Gamma \vdash \phi_1 \bimp \phi_2$, and the premiss is
$\Gamma, \phi_1[e] \vdash \phi_2$ where $e = 1$.
Hence by the induction hypothesis we have that
$\Gamma(\Delta'[0]), \phi_1[e] \vdash \phi_2$ has a closed resource derivation,
and hence so does
$\Gamma(\Delta'[0]) \vdash \phi_1 \bimp \phi_2$.

\item[$*$L:]
In this case, the endsequent is $\Gamma(\phi_1 * \phi_2[e]) \vdash \phi$ where $e = 1$,
and the premiss is  $\Gamma(\phi_1[e], \phi_2[e]) \vdash \phi$.
Hence by the induction hypothesis we have that
$\Gamma(\phi_1[e], \phi_2[e])(\Delta'[0]) \vdash \phi$ has a closed
resource derivation, and hence so does
$\Gamma(\phi_1[e] * \phi_2[e])(\Delta'[0]) \vdash \phi$.

\item[$*$R:]
In this case, the endsequent is $\Gamma \vdash \phi_1 * \phi_2$, and the premisses are
$\Gamma.V \vdash \phi_1$ and  $\Gamma.\ov{V} \vdash \phi_2$.
Hence by the induction hypothesis we have that
$\Gamma.V(\Delta'[0]) \vdash \phi_1$ and $\Gamma.\ov{V}(\Delta'[0]) \vdash \phi_2$
have closed resource derivations, and as $(\Gamma.V)(\Delta'[0]) =
\Gamma(\Delta'[0]).V$, so does
$\Gamma(\Delta'[0]) \vdash \phi_1 * \phi_2$.

\item[$\rightarrow$L:]
In this case, the endsequent is $\Gamma(\phi_1 \rightarrow \phi_2[e]) \vdash \phi$ where
$e = 1$, and the premisses are
$\Gamma \vdash \phi_1$ and $\Gamma(\phi_2[e]) \vdash \phi$.
Hence by the induction hypothesis we have that
$\Gamma(\Delta'[0]) \vdash \phi_1$ and $\Gamma(\phi_2[e])(\Delta'[0]) \vdash \phi$ have
closed resource derivations, and hence so does
$\Gamma(\phi_1 \rightarrow \phi_2[e])(\Delta'[0]) \vdash \phi$.

\end{itemize}
\end{proof}

We are now in a position to show the completeness of resource proofs
for \BI.

\begin{prop}[completeness of resource proofs]\label{prop:bicomplete}
Let $\Gamma \vdash \phi$ be a sequent in \BI.
If $\Gamma \vdash \phi$ has a proof $\Phi$ in {\bf LBI},
then there is a set of Boolean variables
$V$ such that $\Gamma . V \vdash \phi$ has a resource proof
$R$ and the {\bf LBI} proof tree corresponding to $R$ is $\Phi$.
\end{prop}

\begin{proof}

We proceed by induction on the height of the {\bf LBI} proof.

In the base case, the rule used is one of Axiom, $I$R, $1$R and $\perp$L.
We only give the argument for Axiom, the others being similar:

%
In this case, the sequent is just $\phi \vdash \phi$, and it is clear that
this {\bf LBI} proof corresponds to the resource proof
$\ds\frac{}{\phi[x] \vdash \phi}$ with $x = 1$.

%
%

Hence we assume that the result holds for all proofs of no more than a given size.

There are numerous cases, of which we only give the argument for
$W$, $\bimp$L, $\bimp$R, $*$L, $*$R, $\rightarrow$L, $\rightarrow$R, and
$\land$R, the others being similar.

\begin{itemize}


\item[W:]
In this case, the conclusion is $\Gamma(\Delta; \Delta') \vdash \phi$
and the premiss is $\Gamma(\Delta) \vdash \phi$, and so by the
hypothesis there is a resource proof of $\Gamma(\Delta).V \vdash \phi$.
Now as in such a resource proof all the expressions in $\Gamma(\Delta)$ must
be mapped to 1 under the corresponding Boolean assignment, and
$(\Gamma_1, \Gamma_2)[e] = \Gamma_1[e], \Gamma_2[e]$, it is
clear that there is a resource proof of $\Gamma(\Delta; \Delta').V
\vdash \phi$ for which $\Phi$ is the corresponding {\bf LBI} proof.

%
%

\item[$\bimp$L:]
In this case, the conclusion is $\Delta(\Delta', \Gamma, \phi_1 \bimp \phi_2) \vdash \phi$
and the premisses are
$\Gamma \vdash \phi_1$ and $\Delta(\Delta', \phi_2) \vdash \phi$,
and so by the hypothesis there are disjoint sets of variables $V$ and $W$
such that there are resource proofs of
$\Gamma.V \vdash \phi_1$ and $\Delta(\Delta', \phi_2).W \vdash \phi$.
Now, as these are resource proofs, by Lemma \ref{lemma:biweak},
there exists $V'$ such that there are resource proofs of
$(\Gamma, \Delta').V' \vdash \phi_1$ and $\Delta((\Gamma,\Delta').\ov{V'}, \phi_2) \vdash \phi$.
Hence there is a resource proof $R$ of $\Delta(\Gamma, \Delta', \phi_1 \bimp \phi_2) \vdash \phi$
for which $\Phi$ is the corresponding {\bf LBI} proof.

\item[$\bimp$R:]
In this case, the conclusion is $\Gamma \vdash \phi_1 \bimp \phi_2$ and the premiss is
$\Gamma, \phi_1 \vdash \phi_2$, and so by the
hypothesis there is a resource proof of $(\Gamma, \phi_1[e]).V \vdash \phi_2$, and as $e$
must be $1$, we have that there is a resource proof of
$\Gamma.V, \phi_1[1] \vdash \phi_2$, and
hence there is a resource proof $R$ of $\Gamma.V \vdash \phi_1 \bimp \phi_2$ for which
$\Phi$ is the corresponding {\bf LBI} proof.

\item[$*$L:]
In this case, the conclusion is $\Gamma(\phi_1 * \phi_2) \vdash \phi$ and the premiss is
$\Gamma(\phi_1, \phi_2) \vdash \phi$, and so by the
hypothesis there is a resource proof of $\Gamma(\phi_1, \phi_2).V \vdash \phi$, and
hence there is a resource proof $R$ of $\Gamma(\phi_1 * \phi_2).V \vdash \phi$ for which
$\Phi$ is the corresponding {\bf LBI} proof.

\item[$*$R:]
In this case, the conclusion is $\Gamma, \Delta \vdash \phi_1 * \phi_2$,
and the premisses are
$\Gamma \vdash \phi_1$ and $\Delta \vdash \phi_2$,
and so by the hypothesis there are disjoint sets of variables $V$ and $W$
such that there are resource proofs of
$\Gamma.V \vdash \phi_1$ and $\Delta.W \vdash \phi_2$.
Now, as these are resource proofs, by Lemma \ref{lemma:biweak},
there exists $V'$ such that there are resource proofs of
$(\Gamma, \Delta).V' \vdash \phi_1$ and $(\Gamma,\Delta).\ov{V'} \vdash \phi_2$.
Hence there is a resource proof $R$ of $\Gamma, \Delta \vdash \phi_1 * \phi_2$
for which $\Phi$ is the corresponding {\bf LBI} proof.

\item[$\rightarrow$L:]
In this case, the conclusion is $\Gamma(\phi_1 \rightarrow \phi_2) \vdash \phi$ and the premisses are
$\Gamma \vdash \phi_1$ and $\Gamma(\phi_2) \vdash \phi$, and so by the
hypothesis there are resource proofs of
$\Gamma.V \vdash \phi_1$ and $\Gamma(\phi_2[e]).W \vdash \phi$.
Now $e$ must be $1$ under $W$, and as we must have all formul{\ae} in the
premisses mapped to 1, we have that there is a set $V'$ of distinct variables
$V'$ such that  there are resource proofs of
$\Gamma.V' \vdash \phi_1$ and $\Gamma(\phi_2[e]).V' \vdash \phi$,
and hence there is a resource proof $R$ of
$\Gamma(\phi_1 \rightarrow \phi_2).V' \vdash \phi$
for which $\Phi$ is the corresponding {\bf LBI} proof.

\item[$\rightarrow$R:]
In this case, the conclusion is $\Gamma \vdash \phi_1 \rightarrow \phi_2$ and the premiss is
$\Gamma; \phi_1 \vdash \phi_2$, and so by the hypothesis there is a resource proof
of $(\Gamma; \phi_1[e]).V \vdash \phi_2$, and as $e$ must be $1$,
there is a set of distinct variables $V'$ such that there is a resource proof of
$\Gamma.V'; \phi_1 \vdash \phi_2$, and
hence there is a resource proof $R$ of $\Gamma.V' \vdash \phi_1 \rightarrow \phi_2$ for which
$\Phi$ is the corresponding {\bf LBI} proof.


\item[$\land$R:]
In this case, the conclusion is $\Gamma \vdash \phi_1 \land \phi_2$ and the premisses are
$\Gamma \vdash \phi_1$ and $\Gamma \vdash \phi_2$, and so by the
hypothesis there are resource proofs of $\Gamma.V \vdash \phi_1$ and $\Gamma.W \vdash \phi_2$.
Now we must have all formul{\ae} in the premisses mapped to 1, and so
we have that there is a set $V'$ of distinct variables $V'$ such that
there are resource proofs of
$\Gamma.V' \vdash \phi_1$ and $\Gamma.V' \vdash \phi_2$, and
hence there is a resource proof $R$ of $\Gamma.V' \vdash \phi_1 \land \phi_2$ for which
$\Phi$ is the corresponding {\bf LBI} proof.

%

\end{itemize}
\end{proof}

\section{Strategies} \label{sec:strategies}

Hsaving established that the use of Boolean constraints is sound and
complete for {\bf MLL} and two extensions of it, we now turn to the
issue of how the constraints generated may be solved. As mentioned
above, this is independent of the inference rules themselves.

Clearly there are many different search strategies which could be used
to generate a solution to the constraints, but the strategies which we
wish to consider here divides conveniently (but not exhaustively) into
three: \emph{lazy}, \emph{eager} and \emph{intermediate}.  As
mentioned above, resource proofs are intended to be independent of a
particular strategy, but to contain an explicit specification of the
distributive constraints. Strategies are thus distinguished by the
manner of solution of the constraints generated during
proof-search. We denote as an \emph{n-strategy} one which solves the
equations from at most $n$ multiplicative branches at a time.

~\\
\noindent {\bf Lazy distribution.} In terms of the calculus introduced above,
lazy distribution solves one multiplicative branch's worth of Boolean
constraints at a time (thus making it a 1-strategy), and propagates
the solution together with any remaining constraints to the next
multiplicative branch. This may be thought of as a pessimistic
strategy, in that as only a minimal set of constraints is solved, if
the derivation turns out to be unsuccessful, then only a minimal
amount of work has been done. This strategy is the one most commonly
used in linear logic programming languages such as Lygon
\cite{hpw96,wh95} and Lolli \cite{hm94}, and is analogous to depth-first search.

~\\
\noindent {\bf Eager distribution.}
The eager distribution is an $\omega$-strategy, in that an unbounded
number of equations may be solved, and so all leaves must be closed
before any attempt is made to solve the set of constraints. Hence a
constraint solver would be called only once per derivation, with a
potentially large set of constraints. This may be thought of as an
optimistic strategy, in that if one of the branches leads to failure,
then the work done on evaluating all the other branches in parallel
has been wasted.
\footnote{It may also be
useful to check that the current constraints have a solution (as
distinct from actually solving them), as happens in many constraint
logic programming languages.}

Note that, analogous to the differences between depth-first and
breadth-first search, there are examples in which an eager strategy
is preferable to a lazy one. For example, consider the sequent
$\Gamma \vdash p \otimes q$ where $\Gamma \vdash p$ is
provable and $q$ does not occur anywhere in $\Gamma$ (and hence
$\Gamma' \not \vdash q$ for any submultiset $\Gamma'$ of $\Gamma$). Clearly,
by an appropriate choice of $\Gamma$ and $p$, the proof of $\Gamma
\vdash p$ may be made arbitrarily complex (or, for that matter,
infinite if $\Gamma \vdash p$ belongs to an undecidable class of
sequents). A lazy strategy will generate the sequents

$$
\infer{\Gamma \vdash p \otimes q}
{
    \Gamma \vdash p & \vdash q
}
$$

\noindent
and hence spend an arbitrarily large amount of time on the proof of
$\Gamma \vdash p$ (which it will eventually discover is provable) when
it is clear that the other branch will fail immediately. However, an
eager distribution, which attempts to solve all branches in parallel,
will detect that there is no way to close the right-hand branch before
any significant amount of work is done, as there is no way to form a
putative axiom out of this branch (as $q$ does not appear anywhere in
$\Gamma$).  Hence an eager strategy will be more efficient than
a lazy version in this case.

~\\
\noindent {\bf Intermediate distribution.}
Intermediate strategies are $n$-strategies, where $n \geq 2$, and are
analogous to the technique of \emph{iterative deepening}. The precise
way in which a proof which involves $n+1$ multiplicative branches may
be either an eager search for the first $n$ such branches (proceeding
from the root), and then lazy searches from then on (effectively
performing $n$ lazy searches in parallel), or to ``switch'' the eager
version to a place further from the root (effectively performing a
number of lazy searches, one of which is a $n$-way eager search).

As in general it is not possible to predict in advance where the
leaves in a proof will be found, it would seem intuitively reasonable
to adopt the policy that the eager behaviour occurs towards the root,
and once the bound of $n$ is reached, $n$ multiplicative branches are
chosen to be explored in a lazy manner. However, this may result in
sub-optimal behaviour, as the ``locality'' of the constraints is lost.
The alternative would require extra analysis, as initially search
would proceed as above, but once the limit is reached, it is necessary
to re-assign the $n$ searchers to work on the sub-branches of a
particular multiplicative branch in some appropriate way.

For example, consider a 2-strategy with the sequent $p,p,q,q \vdash (p
\otimes q) \otimes (p \otimes q)$. It is easy to see that as there are
3 occurrences of $\otimes$ in the formula in the succedent, there will
be (at least) 4 multiplicative branches in the ensuing proof. Hence it
would be reasonable to use the lazy manner for the first occurrence of
$\otimes$, and then solve each generated branch in an eager manner.

~\\
\noindent{\bf Other strategies.} The three possibilities of lazy, eager and
intermediate are probably the most natural choices of strategies, but they
are not the only ones. One such strategy (which may be thought of as a variation
of a purely lazy strategy) may be described as ``fact-first'', in that
given a sequent such as $\Gamma \vdash p \otimes q$, we can note that
if $q \in \Gamma$ then one of the multiplicative branches will form an
axiom. Hence, rather than arbitrarily selecting one of the branches
(the purely lazy strategy), we select the putative axiom, generate the
appropriate constraints, and continue with the other branch. Thus the
strategy is adaptive, in that the order of evaluation will depend on
the sequents generated.

~\\
\noindent {\bf Examples.} Consider the {\bf MLL} sequent
$p,p,q,q \vdash (p \otimes q) \otimes (p \otimes q)$.

A resource proof of this sequent is of the form

{\small
$$\ds
\infer[\quad{,}]{p,p,q,q \vdash (p \otimes q) \otimes (p \otimes q)}
{
        \infer{p[x_1], p[x_2], q[x_3], q[x_4] \vdash p \otimes q}
        {
                P_1 \qquad
                &
                \qquad P_2
        }
        \qquad
        &
        \qquad
        \infer{p[\ov{x_1}], p[\ov{x_2}], q[\ov{x_3}], q[\ov{x_4}] \vdash p
               \otimes q}
        {
                P_3 \qquad
                &
                \qquad P_4
        }
}
$$
}

\noindent where the leaves are as follows:

{\small
\begin{center}
$P_1:$ $p[x_1.y_1], p[x_2.y_2], q[x_3.y_3], q[x_4.y_4] \vdash p$ \qquad
$P_3:$ $p[\ov{x_1}.z_1], p[\ov{x_2}.z_2], q[\ov{x_3}.z_3], q[\ov{x_4}.z_4]
\vdash p$ \\
$P_2:$ $p[x_1.\ov{y_1}], p[x_2.\ov{y_2}],
q[x_3.\ov{y_3}],q[x_4.\ov{y_4}] \vdash q$ \qquad
$P_4:$ $p[\ov{x_1}.\ov{z_1}], p[\ov{x_2}.\ov{z_2}],
q[\ov{x_3}.\ov{z_3}], q[\ov{x_4}.\ov{z_4}] \vdash q$
\end{center}}

\noindent The lazy strategy yields the following sequence of
constraints and solutions:

\begin{center}
{\small
\begin{tabular}{|cc|cc|c|} \hline
{\bf Leaf} & \qquad & {\bf Constraints added} & \qquad & {\bf
Solutions} \\ \hline \hline
$P_1$      & & $x_1.y_1 = 1, x_2.y_2 = 0, x_3.y_3 = 0, x_4.y_4 = 0$
& & $x_1 = 1, y_1 = 1$ \\ \hline
$P_2$ & & $x_2.\ov{y_2} = 0, x_3.\ov{y_3} = 1, x_4.\ov{y_4} = 0$
& & $x_3 = 1, y_3 = 0, x_2 = 0, x_4 = 0$ \\  \hline
$P_3$ & & $z_2 = 1, z_4 = 0$ & & $z_2 = 1, z_4 = 0$  \\ \hline
$P_4$ & & $\ov{x_4}.\ov{z_4} = 1$ & &  \\ \hline
\end{tabular}}
\end{center}

\noindent which gives us the overall solution

{\small
\begin{center}
${\bf x_1 = 1, x_2 = 0, x_3 = 1, x_4 = 0,}$ \\
${\bf y_1 = 1, y_2 = 0, y_3 = 0, y_4 = 0,}$ \\
${\bf z_1 = 0, z_2 = 1, z_3 = 0, z_4 = 0}$
\end{center}}

\noindent where $y_2, y_4, z_1$ and $z_3$ have been arbitrarily assigned the
value $0$. Note that we can conclude from the equations $x_4.y_4 = 0$ and
$x_4.\ov{y_4} = 0$ that $x_4$ must be $0$, and similarly for $x_2$.

The eager strategy collects the entire set of equations below, and
then solves it to produce the same overall solution.

{\small
\begin{center}
${\bf x_1.y_1 = 1, x_2.y_2 = 0, x_3.y_3 = 0, x_4.y_4 = 0}$ \qquad
${\bf \ov{x_1}.z_1 = 0, \ov{x_2}.z_2 = 1, \ov{x_3}.z_3 = 0,
\ov{x_4}.z_4 = 0}$ \\
${\bf x_1.\ov{y_1} = 0, x_2.\ov{y_2} = 0, x_3.\ov{y_3} = 1,
x_4.\ov{y_4} = 0}$ \qquad
${\bf \ov{x_1}.\ov{z_1} = 0, \ov{x_2}.\ov{z_2} = 0, \ov{x_3}.\ov{z_3} = 0,
\ov{x_4}.\ov{z_4} = 1}$ \\
\end{center}}

One variant of the intermediate strategy first solves the
equations for $P_1$ and $P_3$ in parallel, and then those for $P_2$
and $P_4$:

\begin{center}
{\small
\begin{tabular}{|cc|cc|c|}  \hline
{\bf Leaf} & \qquad & {\bf Constraints added} & \qquad & {\bf Solutions} \\ \hline \hline
$P_1, P_3$ & & $x_1.y_1 = 1, x_2.y_2 = 0, x_3.y_3 = 0, x_4.y_4 = 0$
& & $x_1 = 1, y_1 = 1$ \\ \hline
& & $\ov{x_1}.z_1 = 0, \ov{x_2}.z_2 = 1, \ov{x_3}.z_3 = 0,
\ov{x_4}.z_4 = 0$ & & $x_2 = 0, z_2 = 1$ \\ \hline
$P_2, P_4$ & & $x_2.\ov{y_2} = 0, x_3.\ov{y_3} = 1, x_4.\ov{y_4} = 0$
& & $x_3 = 1, y_3 = 0$ \\  \hline
& & $\ov{x_3}.\ov{z_3} = 0, \ov{x_4}.\ov{z_4} = 1$ & & $x_4 = 0, z_4 = 0$\\ \hline
\end{tabular}}
\end{center}

The other variant of the intermediate strategy first solves the
equations for $P_1$ and $P_2$ in parallel, and then those for $P_3$
and $P_4$:

\begin{center}
{\small
\begin{tabular}{|cc|cc|c|} \hline
{\bf Leaf} & \qquad & {\bf Constraints added} & \qquad & {\bf
Solutions} \\  \hline \hline
$P_1, P_2$ & & $x_1.y_1 = 1, x_2.y_2 = 0, x_3.y_3 = 0, x_4.y_4 = 0$
             & & $x_1 = 1, y_1 = 1$ \\ \hline
& & $x_1.\ov{y_1} = 0, x_2.\ov{y_2} = 0, x_3.\ov{y_3} = 1, x_4.\ov{y_4} = 0$
& & $x_3 = 1, y_3 = 0, x_4 = 0, x_2 = 1$ \\ \hline
$P_3, P_4$ & & $z_2 = 1, z_4 = 0$  & & $z_2 = 1, z_4 = 0$ \\ \hline
        & & $z_2 = 1, \ov{z_4} = 1$  & & \\ \hline
\end{tabular}}
\end{center}

\noindent Note that solving the equations for $P_1$ and $P_2$ in parallel
generates more of the solution than solving those for $P_1$ and $P_3$
in parallel.

Consider now the {\bf PLL} sequent
$p, q, q \vdash (p \otimes q) \oplus (p \otimes q \otimes q)$.

This has the following resource derivation.

$$
\infer[\oplus {\rm R}]{p, q, q \vdash (p \otimes q) \oplus (p \otimes q \otimes q)}
{
    \infer[\otimes {\rm R}]{p, q, q \vdash (p \otimes q)[x], (p \otimes q \otimes q)[\ov{x}]}
    {
        \infer[]{p[y_1], q[y_2], q[y_3] \vdash (p \otimes q)[0], p \otimes q}
        {
                P_1 \qquad
                &
                \qquad P_2
        }
        &
        p[\ov{y_1}], q[\ov{y_2}], q[\ov{y_3}] \vdash (p \otimes q)[0], q
    }
}
$$

where $P_1$ and $P_2$ are as follows:
\begin{center}
$P_1: p[y_1.z_1], q[y_2.z_2], q[y_3.z_3] \vdash (p \otimes q)[0], p$ \\
$P_2: p[y_1.\ov{z_1}], q[y_2.\ov{z_2}], q[y_3.\ov{z_3}] \vdash (p
\otimes q)[0], q$ \\
\end{center}

Solving the equations gives us

\begin{center}
$x = 0, y_1 = 1, y_2 = 1, y_3 = 0, z_1 = 1, z_2 = 0, z_3 = 0$
\end{center}

Note that $x = 0$ because of  the choice of principal formula in the
application of the $\oplus$R rule.

This gives us the resource proof

$$
\infer[\oplus {\rm R}]{p, q, q \vdash (p \otimes q) \oplus (p \otimes q \otimes q)}
{
    \infer[\otimes {\rm R}]{p, q, q \vdash (p \otimes q)[0], (p \otimes q \otimes q)[1]}
    {
        \infer[]{p[1], q[1], q[0] \vdash (p \otimes q)[0], p \otimes q}
        {
                p[1], q[0], q[0] \vdash (p \otimes q)[0], p
                &
                p[0], q[1], q[0] \vdash (p \otimes q)[0], q
        }
        &
        p[0], q[0], q[1] \vdash (p \otimes q)[0], q
    }
}
$$

and hence the sequent proof

$$
\infer[\oplus {\rm R}]{p, q, q \vdash (p \otimes q) \oplus (p \otimes q \otimes q)}
{
    \infer[\otimes {\rm R}]{p, q, q \vdash p \otimes q \otimes q}
    {
        \infer[]{p, q \vdash p \otimes q, p \otimes q}
        {
                p \vdash p
                &
                q \vdash q
        }
        &
        q \vdash q
    }
}
$$

Note that, in this case, the various strategies will only differ on the
sub-proof $p, q, q \vdash p \otimes q \otimes q$ in the manner
discussed above.

Turning to an example of an unprovable sequent, consider

$$
p \otimes q, r \vdash p \otimes q
$$

The only possible resource derivation will look like this:

$$
\infer[\otimes {\rm L}]{p \otimes q, r \vdash p \otimes q}
{
    \infer[]{p, q, r \vdash p \otimes q}
    {
        p[x_1], q[x_2], r[x_3] \vdash p
        &
        p[\ov{x_1}], q[\ov{x_2}], r[\ov{x_3}] \vdash q
    }
}
$$

The lazy strategy would solve, say, the left-hand leaf first,
resulting in the equations

\begin{center}
$x_1 = 1, x_2 = 0, x_3 = 0$
\end{center}

which would make the second leaf into

\begin{center}
$p[0], q[1], r[1] \vdash q$
\end{center}

which clearly cannot be made into an axiom, and the search fails at
this point.

The eager strategy would generate the equations

\begin{center}
$x_1 = 1, x_2 = 0, x_3 = 0$
\end{center}
from the left-hand leaf, together with the equations

\begin{center}
$x_1 = 1, x_2 = 0, x_3 = 1$
\end{center}

from the second leaf, and hence determine that the union of these two
sets of equations has no solution.

Turing to an example from \BI, consider the sequent

$$(r, (p;t), p \bimp q); s \vdash q * r$$

This results in the following resource derivation:

{\small
$$
   \infer[-* {\rm L}]{(r, (p;t), p \bimp q); s \vdash q * r}
    {
      \infer[W]{r[x_1], (p;t)[x_2] \vdash p}{r[x_1], p[x_2] \vdash p}
      &
      \infer[W]{(r[\ov{x_1}], (p;t)[\ov{x_2}], q);s \vdash q * r }
      {
        \infer[* {\rm R}]{r[\ov{x_1}], (p;t)[\ov{x_2}], q \vdash q * r }
        {
           r[\ov{x_1}.y_1], (p;t)[\ov{x_2}.y_2], q[y_3] \vdash q
           &
           r[\ov{x_1}.\ov{y_1}], (p;t)[\ov{x_2}.\ov{y_2}], q[\ov{y_3}] \vdash r
        }
      }
    }
$$
}

Note that the fact that $(r, (p;t), p \bimp q); s$ has an outermost
additive context means that the distribution problem for the $\bimp$L
rule is to determine how to distribute the multiplicative sub-context $r, (p;t)$.

A lazy evaluation using the leftmost leaf first will give us

\begin{center}
$x_1 = 0, x_2 = 1$
\end{center}

Propagating this to the next leaf will give us

$$r[y_1], (p;t)[0], q[y_3] \vdash q $$

from which we determine $y_1 = 0, y_3 = 1$. Passing this onto the third leaf gives us

$$r[1], (p;t)[0], q[0] \vdash r$$

as required.

An eager evaluation would generate the set of equations below.

$$ x_1 = 0, x_2 = 1, \ov{x_1}.y_1 = 0, \ov{x_2}.y_2 = 0, y_3 = 1,
\ov{x_1}.\ov{y_1} = 1, \ov{x_2}.\ov{y_2} = 0, \ov{y_3} = 0 $$

Solving this set of equations gives

$$x_1 = 0, x_2 = 1, y_1 = 0, y_2 = , y_3 = 1$$,

as above.

\section{The generality of the method} \label{sec:general}

We have seen that the method of calculating the distribution of
formul{\ae} during proof-search with multiplicative rules via
solving systems of Boolean constraint equations may be applied
to both linear logic and \BI. Thus these two systems illustrate
the use of our methods for the two systems' different treatments of
Weakening and Contraction, and for the relationship between linear
and intuitionistic implication, based both on exponentials and on
bunching.

Whilst \BI\ can be seen as the free combination of its linear and
intuitionistic fragments, the family of ``traditional'' relevant
logics, as represented in \cite{Read88,Restall2000},  make choices
of connectives and their associated laws which, at least from a semantic
perspective, are very much more \emph{ad hoc}.

However, Read's taxonomy \cite{Read88} provides a setting for a
systematic assessment of the applicability of our methods. Read
identifies a basic substructural system, {\bf DW}, and then considers
a zoo of possible extensions. We conjecture that our method is
sufficiently general to deal with essentially any system involving
multiplicative rules. Below we briefly consider the systematic
classification of a large family of systems of relevant logic
presented by Read \cite{Read88}.

The basic system, {\bf DW}, is formulated as a bunched natural deduction
system \cite{Read88,Pym00Mono}. The collection of connectives is
essentially the same as that taken in \BI\ and the natural deduction
system is, of course, closely related to \LBI. The introduction rules
are the same as the right rules of \LBI . The elimination rules stand in
the usual relationship to the left rules of \LBI. For example, the elimination
rules for the multiplicative (intensional) conjunction are implication are,
respectively,\footnote{We use a natural deduction system here in order to
conveniently characterize the a collection of relevant systems. However,
the failure of the subformula property for rules such as $\lolli\mbox{\rm E}$
makes effective search problematic.}
$$
\otimes\mbox{\rm E}\quad\frac{\Gamma \vdash \phi \otimes \phi \quad
\Delta(\phi,\psi) \vdash \chi}
{\Delta(\Gamma) \vdash \chi}
\quad\mbox{\rm and}\quad
\lolli\mbox{\rm E}\quad\frac{\Gamma \vdash \phi \lolli \psi \quad \Delta \vdash \phi}
{\Gamma , \Delta \vdash \psi}.
$$

The exceptions are the rules for (a classical) negation.
The rules for negation are, in our notation,\footnote{We use ``,'' for
multiplicative bunching and ``;'' for additive bunching; Read's notation
is the exact opposite. We write $\emptyset_{m}$ for the multiplicative
identity; Read uses t.}
$$
\sim\mbox{\rm I}\quad\frac{\Gamma , \phi \,\vdash\, \sim\psi \quad \Delta \,\vdash\, \phi}
{\Gamma , \Delta \,\vdash\, \sim \phi}
\qquad\mbox{\rm and}\qquad
\sim\mbox{\rm E}\quad\frac{\Gamma \,\vdash\, \sim\sim \phi}{\Gamma \,\vdash\, \phi},
$$
with the optional additional rule,
$$
\mbox{\rm CM}\quad\frac{\Gamma , \phi \,\vdash\, \sim \phi}{\Gamma \,\vdash\, \sim \phi}\quad
\mbox{``consequentia mirabilis''.}
$$
The structural rules of {\bf DW} are:
\begin{itemize}
\item The monoid (identity, commutativity, associativity) laws for additive bunching;
\item Weakening and Contraction for additive bunching;
\item A left-identity for multiplicative bunching,
$\emptyset_m , \Gamma \equiv \Gamma$.
\end{itemize}
Note that the basic system {\bf DW} does not assume commutativity of ``,''.

To describe the various additional structural properties which a system, we
write $\Gamma \leq \Gamma'$, for bunches $\Gamma$, $\Gamma'$ to denote that
$\Delta(\Gamma) \vdash \phi$ implies $\Delta(\Gamma') \vdash \phi$.
The family of relevant logics can then be presented systematically
as a hierarchy of systems regulated by the axioms given, using Read's
terminology but, as above, our notation, in
Table~\ref{table:relevant-axioms}.

\begin{table}
\hspace{5mm}
\begin{center}
\begin{tabular}{|c|c|c|} \hline
B        &  $\Gamma , (\Delta , \Theta) \leq (\Gamma , \Delta) , \Theta$  &  prefixing                   \\ \hline
B$'$     &  $\Gamma , (\Delta , \Theta) \leq (\Delta , \Gamma) , \Theta$  &  suffixing                   \\ \hline
C$^{**}$ &  $\Gamma , \emptyset_{m} \leq \Gamma$                          &  right-identity              \\ \hline
C$^{*}$  &  $\Gamma , \Delta \leq \Delta , \Gamma$                        &  assertion                   \\ \hline
C        &  $(\Gamma , \Delta) , \Theta \leq (\Gamma , \Theta) , \Delta$  &  permutation                 \\ \hline
W$^{*}$  &  $\Gamma  , \Gamma \leq \Gamma$                                &  conjunctive assertion       \\ \hline
W        &  $(\Gamma , \Delta) , \Delta$                                  &  contraction                 \\ \hline
WB       &  $\Gamma , (\Gamma , \Delta) \leq \Gamma , \Delta$             &  conjunctive syllogism       \\ \hline
S$^{*}$  &  $\Delta , (\Gamma , \Delta) \leq \Gamma , \Delta$             &                              \\ \hline
S        &  $(\Gamma , \Theta) , (\Delta,\Theta) \leq (\Gamma , \Delta) , \Theta$ &                      \\ \hline
K$^{*}$  &  $\Gamma \leq \Gamma , \Gamma$                                 &  mingle / premiss repetition \\ \hline
K        &  $\Gamma \leq \Gamma , \Delta$                                 &  affinity                    \\ \hline
\end{tabular}
\end{center}
\hspace{5mm}
\caption{Relevant Axioms}
\label{table:relevant-axioms}
\vspace{0.1in}
\end{table}

We can now define, following \cite{Read88}, the following systems:

$$
\begin{array}{ll}
{\bf TW} = \mbox{\rm B + B$'$}                                 &
           \mbox{\rm {\bf DL} = {\bf DW} + CM + WB}                           \\
{\bf RW} = \mbox{\rm {\bf TW} + C}                             &
           \mbox{\rm {\bf TL} = {\bf TW} + CM + WB (= {\bf DL} + B + B$'$)}   \\
{\bf T}  = \mbox{\rm {\bf TL} + W}                             &
           \mbox{\rm {\bf E}  = {\bf T}  + C$^{**}$}                          \\
{\bf R}  = \mbox{\rm {\bf E} + C$^{*}$ \quad (= {\bf RW} + W)} &
           \mbox{\rm {\bf RM} = {\bf R}  + K$^{*}$}                           \\
\end{array}
$$
\BI\ is the system which takes just identity, associativity and commutativity
for ``,''.

The system {\bf DW}, and so each of the systems defined above, supports the
distributive law for additive conjunction and disjunction:
$$
\phi \wedge (\psi \vee \chi) \dashv\vdash (\phi \wedge \psi) \vee (\phi \wedge \chi).
$$
In this sense, the family of relevant systems is closer to \BI\ than to linear logic.
Indeed, linear logic without the exponentials may be seen, relative to this hierarchy,
as sitting below even {\bf DW}, the rules for its additives not even permitting
distribution. Classical logic
arises as
${\bf R} + {\rm K}$,
so that we recover a connection, via the exponentials,
with the non-distributive, linear systems.

Turning at last to resource proofs, we can make the following observations:
\begin{itemize}
\item The natural deduction-style operational rules of {\bf DW} are,
apart from negation, all handled just as for their counterparts in
\LBI, the introduction rules being the same as the right rules and
the elimination rules typically being simpler than the corresponding
left rules. For example, the resource versions of the $\otimes$E and
$\lolli$E rules may be formulated as, respectively,
$$
\otimes\mbox{\rm E}\quad\frac{\Gamma.V \vdash \phi \otimes \phi \quad
\Delta(\Gamma.\ov{V} , \phi,\psi) \vdash \chi}{\Delta(\Gamma) \vdash \chi}
\quad\mbox{\rm and}\quad
\lolli\mbox{\rm E}\quad\frac{\Gamma.V \vdash \phi \lolli \psi \quad
\Gamma.\ov{V} \vdash \phi}{\Gamma \vdash \psi}.
$$
The other operational rules, excluding negation, are treated similarly;
\item The $\sim$E rule, being unary with only right-hand activity, is trivial.
The resource version of the $\sim$I rule is
$$
\sim{I}\quad \frac{\Gamma.V , \phi[e] \,\vdash\, \sim\psi \quad
\Gamma.\ov{V} \,\vdash\, \phi}
{\Gamma \,\vdash\, \sim \phi}\quad{e=1}.
$$
The CM rule goes as
$$
\mbox{\rm CM}\quad\frac{\Gamma , \phi[e] \vdash \sim\phi}
{\Gamma \vdash \sim\phi}\quad{e=1};
$$
\item The various structural rules can now be seen to be unproblematic. For example,
W$^{*}$, W, WB, S$^{*}$ and S all follow the pattern of Contraction in \BI.
Similarly, K$^{*}$ and K follow the pattern of Weakening in \BI.
\end{itemize}
Our usual soundness and completeness properties can now established for the family
of relevant systems .

Thus our analysis encompasses the full range of established (propositional)
substructural logics. We conjecture that our methods can be applied to the
multiplicative (or intensional) quantifiers introduced in \cite{OP99,pym99lics}
and developed in \cite{Pym00Mono}. The usual additive (or extensional) quantifiers
were treated in \cite{hp97} (they are unproblematic).

\medskip

\noindent {\bf Acknowledgements.} We thank John Crossley, Samin Ishtiaq,
Michael Winikoff and anonymous referees from CADE-14 for their comments
on this work. Partial support from the British Council, the UK EPSRC and
the RMIT Faculty of Applied Science is gratefully acknowledged. Harland
is grateful for the hospitality of the Department of Computer Science of
Queen Mary and Westfield College, University of London during a period of
sabbatical leave. Pym is grateful for the partial support of the EPSRC.


\end{document}